\documentclass{ar2e1}
\usepackage[margin=1in]{geometry}
\usepackage{ARAstroBib,amsmath,amssymb,graphicx}
\def\lesssim{\mathrel{\hbox{\rlap{\hbox{\lower4pt\hbox{$\sim$}}}\hbox{$<$}}}}
\def\gtrsim{\mathrel{\hbox{\rlap{\hbox{\lower4pt\hbox{$\sim$}}}\hbox{$>$}}}}

\begin{document}
\input epsf.tex    

\input psfig.sty

\jname{Author draft of article published as ARA\&A, 49, 195-236 (2011)}

\title{Dynamics of Protoplanetary Disks}

\markboth{Philip J. Armitage}{Dynamics of Protoplanetary Disks}

\author{Philip J. Armitage
\affiliation{JILA, 440 UCB, University of Colorado, Boulder CO80309-0440, USA}}


\begin{abstract}
Protoplanetary disks are quasi-steady structures whose evolution and dispersal 
determine the environment for planet formation. I  
review the theory of protoplanetary disk evolution and its connection to 
observations. Substantial progress has been made in elucidating the physics 
of potential angular momentum transport processes -- including self-gravity, 
magnetorotational instability, baroclinic instabilities, and magnetic 
braking -- and in developing testable models for disk dispersal via photoevaporation. The 
relative importance of these processes depends upon the initial mass, size and 
magnetization of the disk, and subsequently on its opacity, ionization 
state, and external irradiation. Disk dynamics is therefore coupled to 
star formation, pre-main-sequence stellar evolution, and dust coagulation 
during the early stages of planet formation, 
and may vary dramatically from star to star.  The importance of validating 
theoretical models is emphasized, with the 
key observations being those that probe disk structure on the scales, 
between 1~AU and 10~AU, where theory is most uncertain.

\smallskip

Figures and illustrations from this article are available at {\tt http://jila.colorado.edu/$\sim$pja/araa.html}
\end{abstract}

\maketitle

\section{INTRODUCTION}
Protoplanetary disks are the observational manifestation of the initial 
conditions for planet formation. They can be defined as rotationally 
supported structures of gas (invariably containing dust) that surround 
young, normally pre-main-sequence stars. Although most observed 
disks have inferred masses that are a small fraction of the stellar 
mass, no meaningful distinction can be drawn between physical processes in 
protoplanetary disks and those that occur in the earlier, ``protostellar" phase, 
in which both star and disk are accreting rapidly. Similarly, a common set of 
processes operate, albeit to varying degrees, in disks around 
brown dwarfs, Classical T~Tauri stars (low-mass 
pre-main-sequence stars that are actively accreting), and massive stars. 
A clear demarcation does 
separate protoplanetary disks from debris disks; dusty gas-poor structures 
around older stars whose properties reflect the collisional evolution of 
a population of small bodies \citep{wyatt08}. 

Around low-mass stars, protoplanetary disks are persistent; the typical 
lifetime of $\sim 10^6$~years \citep{haisch01} equates to thousands of 
dynamical times at 100~AU. Evolution can only rarely be observed in 
individual objects, and must instead be discerned from statistical 
studies of populations. This slow evolution is a 
consequence of angular momentum conservation; protoplanetary disks 
are nearly stable fluid configurations that evolve under the action 
of relatively slow processes -- angular momentum transport, mass 
infall, and disk winds. These agents control both the secular 
evolution and the eruptive behavior observed in protoplanetary disks, 
and set the overall environment within which planet formation occurs: 
the location of the snow line, the mass and time available to form 
gas giants, the rate of migration, and so forth. Moreover, in most 
models the evolution of the disk is linked to the presence of 
turbulence within the disk gas, and the strength and nature of that 
turbulence controls the growth of solids during 
the earliest phases of planet formation \citep[for a review, see][]{armitage10}.

A generic model for the evolution of protoplanetary disks, based 
on the theory of geometrically thin accretion disks \citep{lyndenbell74}, 
has long been available, but this model lacks predictive power absent 
a detailed specification of the angular momentum transport and mass loss 
processes at work. Pinning these down has taken a long time, but 
substantial progress has now been made in understanding candidate  
angular momentum transport and mass loss mechanisms. This review 
addresses the physics of these mechanisms, how they may combine 
to yield some of the phenomena known to occur in disks around young stars, 
and how they relate to properties of disks that may be observable 
in the future. The focus is theoretical, observational aspects of 
disk structure and evolution are discussed in the review by 
\citet{williams11} in this volume.

\section{A FRAMEWORK FOR DISK EVOLUTION}
Protoplanetary disks are observed to be geometrically thin, in the sense 
that the vertical scale height $h \ll r$, and typically have inferred masses 
$M_{\rm disk} \ll M_*$. These properties imply that models of disk 
evolution should be grounded within the theory of thin accretion 
disks \citep{pringle81}, within which radial pressure 
gradients are negligible, the angular velocity of the fluid is that of a particle 
in a point mass potential, $\Omega_{\rm K} = \sqrt{GM_* / r^3}$, and the specific 
angular momentum is an increasing function of radius, $l \propto \sqrt{r}$.
Accretion then requires that the accreting gas must lose angular momentum, 
and the central problem of thin disk theory is to determine why this should 
occur. For protoplanetary disks, there are two qualitatively distinct 
possibilities; angular momentum may be redistributed within the disk 
(``viscous" disk models, which for reasons of space are the exclusive 
focus of this review) or it may be lost to an external sink (``magnetic 
wind" or ``magnetic braking" models). 

\subsection{Accretion Disk Evolution}
For a thin disk, the hydrostatic vertical structure largely decouples from 
the equations describing the radial evolution, which can be simplified 
by integrating over the vertical thickness \citep[a clear textbook treatment is 
given by][]{frank02}. In this limit, the surface density $\Sigma(r,t)$ of an axisymmetric 
planar disk in a Keplerian potential satisfies,
\begin{equation}
 \frac{\partial \Sigma}{\partial t} =
 \frac{3}{r} \frac{\partial}{\partial r} 
 \left[ 
  r^{1/2} \frac{\partial}{\partial r} 
  \left( 
   \nu \Sigma r^{1/2}
  \right)
 \right], 
\label{eq:2_sigma} 
\end{equation}  
provided that external torques and mass loss can be neglected. 
This equation follows from the conservation of mass and angular momentum, 
and is exactly true for a viscous fluid, where $\nu$ would be the kinematic 
viscosity. 
Under normal conditions (unless $\partial (\nu \Sigma) / \partial \Sigma < 0$) 
solutions to the equation show that gas in the disk will accrete onto the star while 
simultaneously spreading diffusively to large radii. 
Less obviously, it is also valid if $\nu$ 
is an effective viscosity arising from turbulence within the disk, provided 
that the turbulence can be described as a local process \citep{balbus99}. 
Likewise, either a microscopic viscosity or local turbulence results in 
a rate of energy dissipation per unit surface area of the disk,
\begin{equation}
 Q_+ = \frac{9}{8} \nu \Sigma \Omega_{\rm K}^2,
\label{eq:2_qplus} 
\end{equation} 
where $\Omega_{\rm K}$ is the Keplerian angular velocity. At radii where the 
disk can be considered to be in a steady-state, these relations immediately 
specify the variation of $\Sigma$ and effective temperature $T_{\rm eff}$ with 
radius (ignoring for now stellar irradiation),
\begin{eqnarray}
 \nu \Sigma & = & \frac{\dot{M}}{3 \pi} 
 \left[ 
  1 - \sqrt{\frac{r_{\rm in}}{r}} 
 \right] \label{eq:2_nusigma} \\
 T_{\rm eff}^4 & = & 
 \frac{3 G M_* \dot{M}}{8 \pi \sigma r^3} 
 \left[ 
     1 - \sqrt{\frac{r_{\rm in}}{r}} 
 \right].
\end{eqnarray}
Here, $\dot{M}$ is the accretion rate, $\sigma$ the Stefan-Boltzmann 
constant, and we have assumed vanishing torque at an inner radius 
$r_{\rm in}$.

Up to this point, we have gotten away with only very limited assumptions (that the 
turbulence is local, and that there are no external torques), but in return we have only 
predicted one quantity, the effective temperature profile of a steady disk. To 
proceed further, we must specify $\nu$. Here, there are two possibilities. The 
classical approach, which still informs the language and much of the work 
oriented toward the observational modeling of disks, is to postulate a 
simple scaling relation between $\nu$ and some locally defined property of the 
flow. Almost invariably, one takes \citep{shakura73},
\begin{equation}
 \nu = \alpha \frac{c_s^2}{\Omega_{\rm K}} = \alpha c_s h,
\end{equation}
where $c_s$ is the disk sound speed and the second equality follows from 
vertical hydrostatic equilibrium 
($h \simeq c_s / \Omega_{\rm K}$). The parameter $\alpha$ is a dimensionless function 
that will be a constant if the assumed scaling law is obeyed. A variety 
of simple arguments can be used to motivate this scaling, but at heart it is a guess whose 
fidelity must be checked through observations and more detailed 
calculations. Its appeal lies in its simplicity; if we assume that $\alpha$ is 
a constant then we can readily construct a model for both the thermal 
structure (via Equation~\ref{eq:2_qplus}) and evolution (Equation~\ref{eq:2_sigma}) 
of the protoplanetary disk, that has just one free parameter.

A more predictive approach is to try and 
compute the effective viscosity that arises from some turbulent process 
directly. (Often, though somewhat confusingly, the results of such calculations 
are reported in terms of an ``effective $\alpha$".) In the case of disk self-gravity, 
this can sometimes be accomplished via an analytic thermal equilibrium argument, to be 
discussed in \S\ref{sec:self_grav}. More typically, however, there is no 
alternative but to compute the non-linear structure of the turbulence and 
evaluate the resultant stresses. Working in cylindrical polar co-ordinates 
$(r,\phi,z)$, the dominant components of the stress tensor can be 
written as a density-weighted average of the fluctuating 
velocity, ${\bf v}$, and magnetic fields, $\bf B$ \citep{balbus98},
\begin{equation}
 W_{r \phi} = 
  \left\langle 
   \delta v_r \delta v_\phi - \frac{B_r B_\phi}{4 \pi \rho} 
  \right\rangle_\rho,
 \label{eq:2_stresses} 
\end{equation} 
where $\delta v_\phi = v_\phi - r \Omega_{\rm K}$. Written in this form, we see that the 
 stress is the sum of a fluid component, described as the Reynolds stress, 
 and a magnetic or Maxwell stress. Both components are 
 proportional to the energy density in their respective turbulent fields. The 
 $\alpha$ prescription, in this notation, amounts to writing,
 \begin{equation}
  W_{r \phi} = \alpha c_s^2,
\end{equation}
from which we see that $\alpha$ will be a constant if the correlated 
velocity fluctuations, and the Alfv\'en speeds associated with the 
appropriate components of magnetic field, both scale with $c_s$.  

\begin{figure}[t!]
\includegraphics[width=\textwidth]{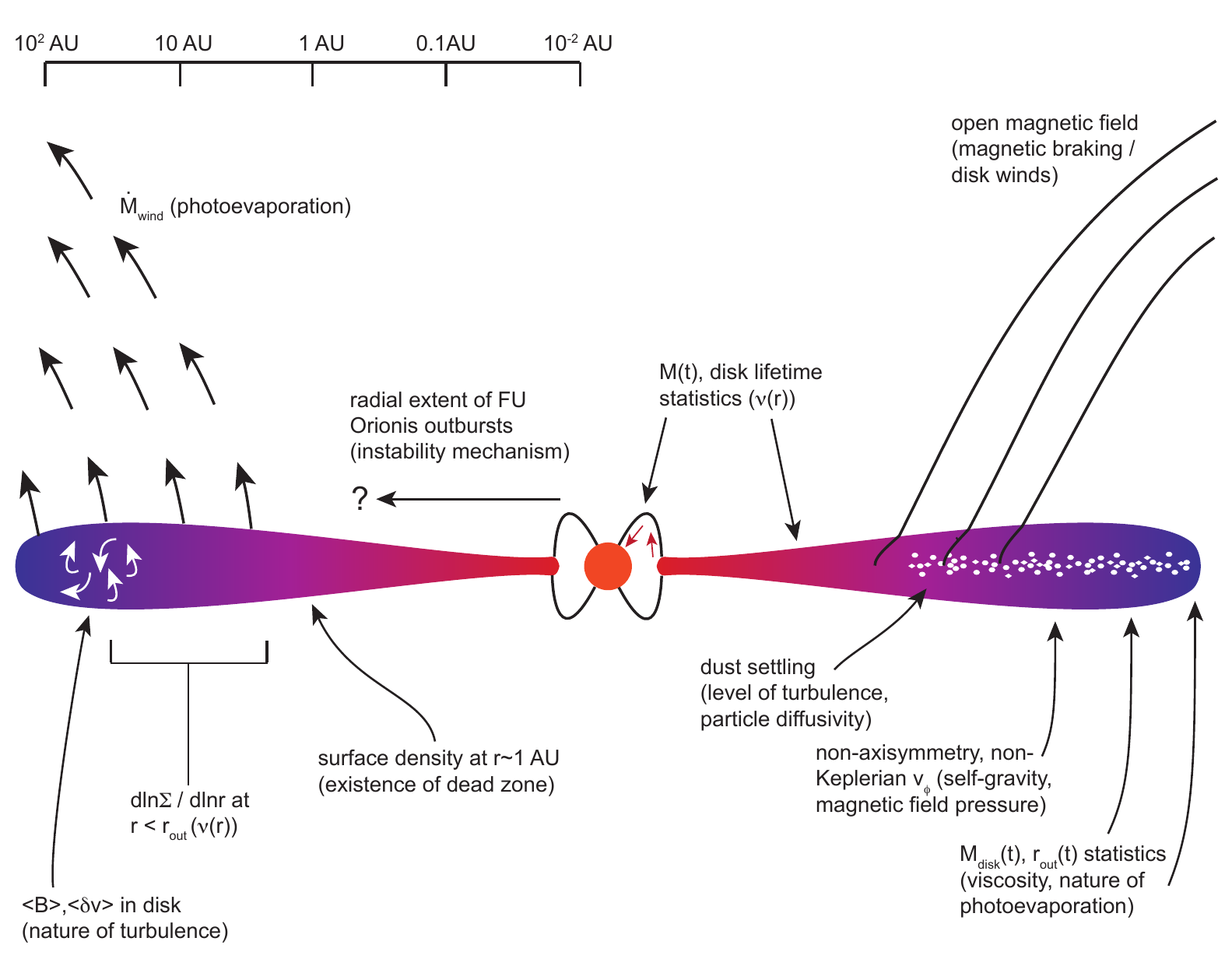}
\caption{Observable diagnostics of protoplanetary disk dynamics. Measurements 
of the local strength and coherence of magnetic fields yield constraints on the relative 
importance of internal versus external angular momentum transport, and on 
the nature of turbulence. The global surface density profile, at radii 
where the disk has attained a steady-state, reflects the radial dependence of 
the efficiency of angular momentum transport. The distribution of quantities 
such as the disk lifetime, $\dot{M}$, $M_{\rm disk}$, and $r_{\rm out}$, 
depends primarily upon the average transport efficiency and the rate 
of wind mass loss. Concordance between multiple independent diagnostics 
tests the overall paradigm for evolution.}
\label{fig:2_observables}
\end{figure}

Although there may be a class of astrophysical disk systems in which the constant 
$\alpha$ approximation works reasonably well, protoplanetary disks are 
probably not a member. We will elaborate on the reasons for this later, 
but for now it suffices to note that magnetic fields, which probably play a crucial 
role in transport (Equation~\ref{eq:2_stresses}), couple to the hot gas in 
the inner disk quite differently than to the cool or diffuse matter further out. 
It would be surprising if this did not result in substantial variations in $\alpha$.
Observationally, this complexity highlights the need to 
obtain multiple constraints on the dynamical processes going on within 
disks, ideally across as wide a range of radii as possible. Many phenomena, 
some of which are illustrated in Figure~\ref{fig:2_observables}, offer 
potential windows into different aspects of the underlying dynamics,
\begin{enumerate}
\item
Measurements of disk magnetic fields, or of turbulent velocity 
fields within the disk \citep{hughes09,hughes10}, yield direct access to the nature of disk turbulence, which is 
connected to the stress (Equation~\ref{eq:2_stresses}). Dust settling, 
and radial mixing of gas or particles, also depend upon the turbulence, but 
these phenomena are less directly related to the stress.
\item
Determinations of the surface density profile \citep[e.g.][]{andrews09}, or joint measurements of 
$\Sigma$ and $\dot{M}$, allow the recovery of $\nu(r)$ 
if the disk is in a steady-state (Equation~\ref{eq:2_nusigma}).
\item
Studies of the evolution of disk populations, for example determination 
of $\langle \dot{M} (t) \rangle$ from observations of clusters of 
different ages \citep[e.g.][]{hartmann98}, constrain angular momentum transport via the 
time dependence implied by Equation~(\ref{eq:2_sigma}). 
\end{enumerate}
Within the basic framework of viscous disk theory, these disparate 
observables constrain a 
single underlying driver of evolution: the stress as measured (say) 
by $\alpha(r,t)$. Demonstrated concordance (or discrepancies) between 
independent measures of $\alpha$ thus represents an empirical route to 
checking the basic assumptions of disk theory.

As \citet{balbus99} have emphasized, the assumption that disks evolve due 
to local turbulent transport is highly restrictive, because in this regime the 
stress uniquely determines the energy dissipation rate as well as the 
angular momentum flux. This need not be true. Angular momentum 
can be redistributed within disks via laminar magnetic torques or 
by global waves, and can be lost entirely if the disk is threaded by 
an open magnetic field that exerts a torque on the disk surface. 
For an open field with a vertical component $B_z$, and an azimuthal 
component $B_{\phi,s}$ (measured at the disk surface), the ratio 
of the external torque to internal Maxwell stresses is \citep[e.g.][]{konigl10},
\begin{equation}
 \frac{{\mathcal T}_{\rm ext}}{{\mathcal T}_{\rm visc}} \sim 
 \frac{B_z B_{\phi,s}}{\langle B_r B_\phi \rangle} 
 \left( \frac{h}{r} \right)^{-1}.
\end{equation} 
Since $(h/r)^{-1} \sim 10$, open magnetic fields can be modestly weaker 
than turbulent fields within the disk and yet drive significant evolution. 
The challenge for ``pure" disk wind and magnetic braking models (those 
which assume that internal torques are unnecessary) is to determine 
whether even weak open fields thread disks across a sufficiently wide range 
of radii, and whether angular momentum loss can occur steadily enough to be 
consistent with the long lifetime of many observed disks. Of course it is 
possible to envisage hybrid models in which external torques dominate 
only in some fraction of cases (perhaps when a collapsing cloud has a 
particularly strong field), or models in which disk winds are only present 
at some radii.

\subsection{$\alpha$ Models of Protoplanetary Disks}
Absent detailed knowledge of how the efficiency of angular momentum transport 
scales with the local physical conditions, the simplest approach to constructing 
viscous disk models is to assume that $\alpha$ is a constant. Computing an 
``$\alpha$ disk model" of this type requires calculating the vertical disk 
structure as a function of $\Omega_{\rm K}$, $\Sigma$ and $\alpha$, taking account of 
the opacity, the mechanism of energy transport between the mid-plane and 
the photosphere, and any incident external radiation flux. Detailed $\alpha$ 
disk models have been constructed by \citet{bell97}, by \citet{dalessio98}, 
and by others. Models in this class are broadly consistent with both imaging and 
Spectral Energy Distribution (SED) data for young stars \citep[e.g.][]{dalessio01}, 
with the main hinderance to a precise test being uncertainties in the 
evolution of the dust within the disk that furnishes the dominant opacity. 
The value of $\alpha$ itself can be constrained by studies of the 
evolution of the stellar accretion rate \citep{hartmann98}, or by 
detailed studies of individual systems \citep{hueso05}. These 
methods typically yield $\alpha \sim 10^{-2}$, with large 
uncertainties. The constraint on $\alpha$ is comparable to 
the elementary estimate that can be derived by setting the global evolution time, 
\begin{equation}
 t_\nu \simeq \frac{r^2}{\nu} = \frac{1}{\alpha \Omega_{\rm K}} \left( 
 \frac{h}{r} \right)^{-2},
\end{equation} 
equal to 1~Myr at a typical disk scale of 30~AU, in a disk with 
$h/r=0.05$ . For planet formation, 
a significant prediction of $\alpha$ models is that 
the steady-state surface density profile 
is roughly $\Sigma \propto r^{-1}$ between a few~AU and $\sim 10^2$~AU, 
significantly flatter than the Minimum Mass Solar Nebula profile $\Sigma \propto 
r^{-3/2}$ \citep{weidenschilling77}. 

\begin{figure}[t!]
\includegraphics[width=\textwidth]{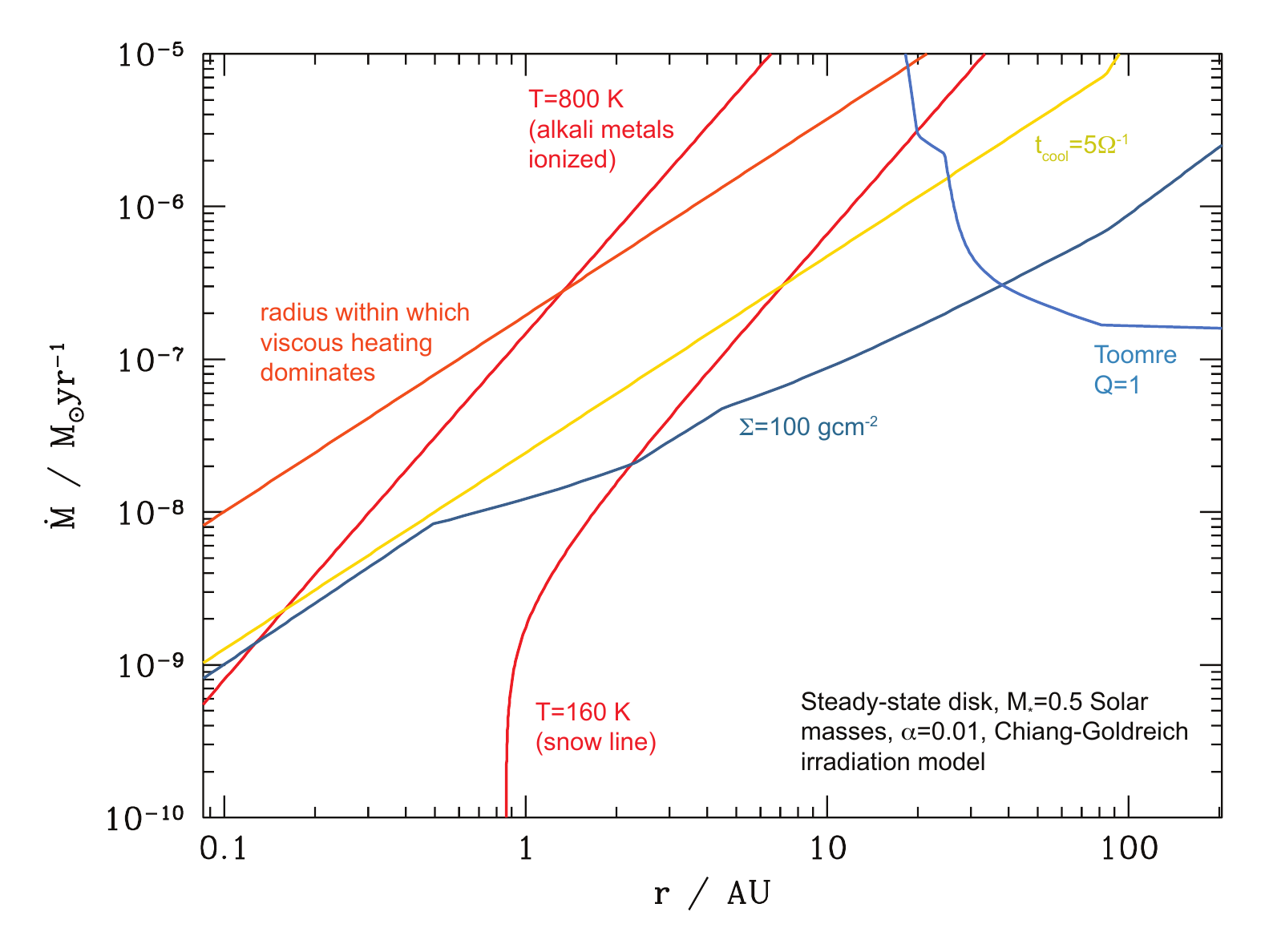}
\caption{Illustration of how critical disk radii, delimiting different physical regimes 
relevant to angular momentum transport and planet formation, scale with accretion rate 
$\dot{M}$. Temperatures are mid-plane values, the cooling time scale is 
defined as $t_{\rm cool} = \Sigma c_s^2 / 2 \sigma T_{\rm eff}^4$, and the 
Toomre $Q$ parameter $Q=c_s \Omega_{\rm K} / \pi G \Sigma$. 
The plot is based upon a one-zone vertical model \citep[calculated as 
described, e.g., by][]{frank02} of a steady-state 
$\alpha$ disk around a $0.5 \ M_\odot$ star, with $\alpha = 10^{-2}$. The opacity 
includes contributions from water ice, amorphous carbon, silicates and graphite (Z. Zhu, 
private communication). 
The calculation is approximate: the radial dependence 
of stellar irradiation is assumed to follow the \citet{chiang97} form, independent 
of both time and disk accretion rate.}
\label{fig:2_conditions}
\end{figure}

Although $\alpha$ models provide a 
parsimonious description of a substantial body of 
data in terms of a single free parameter, there are compelling reasons 
to think that they are not a full description of protoplanetary disks. 
Put simply, no known angular momentum transport mechanism yields 
a constant value of $\alpha$ under protoplanetary disk conditions. 
Transport by magnetohydrodynamic (MHD) processes, for example, is 
predicted to vary in efficacy with the 
degree of ionization, which in turn is a function of the temperature and surface density. 
Similarly, transport by self-gravity, and by possible instabilities that generate 
vortices, scales with the disk cooling time ($t_{\rm cool}$, the ratio of the 
disk thermal energy per unit area to the total flux). 
It is prudent to approach observationally untested $\alpha$ model 
predictions with caution.

Despite these realities, simple $\alpha$ models continue to represent 
a useful starting point for disk studies, since they allow 
a first estimate of how critical quantities such as the ionization 
rate ought to vary with radius in the disk. In this 
spirit, Figure~\ref{fig:2_conditions} shows how several 
significant radii, relevant to understanding where different angular 
momentum transport mechanisms may be effective, scale with 
stellar accretion rate. Although the model used to compute the 
figure is highly simplified, it exposes several generic features 
of protoplanetary disks. First, stellar irradiation provides the 
dominant energy source at large radii, whereas for all but the 
lowest $\dot{M}$ viscous heating dominates close to the 
star \citep{dalessio98}. This will be true unless the disk 
geometry is such as to throw the outer regions into shadow. 
Second, there is a region at $r \sim 1$~AU where the disk is 
cool and dense, and thus likely to have a very low ionization 
fraction. Careful consideration is needed to assess whether 
MHD processes can provide angular momentum transport 
in this region. Third, the Toomre $Q$ parameter, which describes 
the linear stability of a disk to gravitational instability \citep{toomre64}, 
can be low enough for instability to develop at large radii for 
accretion rates above $\sim 10^{-7} \ M_\odot \ {\rm yr}^{-1}$. 
Finally, the cooling time scale is initially (when $\dot{M}$ is 
high) large interior to $r \sim 10^2$~AU, but the line of 
dynamical time scale cooling sweeps inward to small radii 
as $\dot{M}$ drops and stellar irradiation increasingly dominates 
the energy budget of the disk.

\section{ANGULAR MOMENTUM TRANSPORT}
The most intensively studied mechanisms for angular momentum transport 
within viscous protoplanetary disks are self-gravity and the magnetorotational 
(or Balbus-Hawley) instability. These originate from linear instabilities. Hydrodynamic 
instabilities associated with the disk's radial thermal structure 
(``baroclinic instability") could also contribute to angular momentum 
transport, through the non-linear formation of vortices. 
In this Section, we review the current 
understanding these processes. We also discuss, briefly, convection and angular momentum 
transport catalyzed by embedded planets, which might be important 
locally in a subset of protoplanetary disks. 
For want of space the alternate possibility -- that angular momentum loss due 
to magnetic braking is the dominant evolutionary process -- is not 
discussed; the reader seeking details of these models is referred to 
the comprehensive review by \citet{konigl10}.

\subsection{Self-gravity}
\label{sec:self_grav}

The fact that the self-gravity of a massive disk, subject to radiative cooling, 
would result in angular momentum transport and accretion, was recognized 
by \citet{paczynski78}. 
In a non-magnetized disk, the stability of over-dense perturbations is 
set by the balance between self-gravity, pressure (which stabilizes small 
scales against collapse), and shear (which stabilizes large scales). The 
linear stability of an initially axisymmetric thin disk, rotating with 
angular velocity $\Omega (r)$, is quantified by the Toomre $Q$ 
parameter \citep{toomre64},
\begin{equation}
 Q \equiv \frac{c_s \kappa}{\pi G \Sigma},
\label{eq:2_toomre} 
\end{equation}
where $\kappa$, the epicyclic frequency, is the frequency at which 
a radially displaced particle oscillates. Instability to the growth of axisymmetric  
perturbations occurs for $Q \lesssim 1$, though more general disturbances 
\citep{papaloizou91} can grow whenever $Q \lesssim 1.5$. 
In the absence of an external potential, 
the epicyclic frequency for a low mass disk ($M_{\rm disk} \ll M_*$) 
can be written as $\kappa^2 \equiv (2 \Omega / r) 
{\rm d} / {\rm d}r (r^2 \Omega) \simeq \Omega_{\rm K}^2$. This simplification is often used 
for protoplanetary disks.
 If the gas has mean molecular weight 
$\mu$, attaining $Q=1$ requires that the surface density satisfy,
\begin{equation}
 \Sigma > 7.9 \times 10^2 
 \left( \frac{\mu}{2.4} \right)^{-1/2}
 \left( \frac{M_*}{M_\odot} \right)^{1/2}
 \left( \frac{r}{10 \ {\rm AU}} \right)^{-3/2}
 \left( \frac{T_c}{20 \ {\rm K}} \right)^{1/2} \ {\rm g \ cm}^{-2},
\end{equation}
where $T_c$ is the mid-plane temperature. To express this 
more intuitively, we note that the mass enclosed within radius 
$r$ is, approximately, $M_{\rm disk} (r) \sim \pi r^2 \Sigma$. 
Noting that $h = c_s / \Omega_{\rm K}$, an equivalent but approximate 
local condition for instability is,
\begin{equation}
 \frac{M_{\rm disk} (r)}{M_*} \gtrsim \frac{h}{r}.
\end{equation}
Gravitational instability is thus likely to occur early, when protoplanetary 
disks are still massive, and at large radii. Indeed, if irradiation can be 
ignored (usually, it cannot), a steady-state $\alpha$ disk will always 
become Toomre unstable beyond some critical radius. Once a 
disk becomes gravitationally unstable, numerical simulations 
are required in order to assess the subsequent evolution \citep{laughlin94}. 
Three broad classes of outcomes have been identified. The disk 
may settle into a quasi-steady, ``saturated" state of self-gravitating 
turbulence, in which trailing spiral arms yield an outward transport 
of angular momentum via gravitational torques \citep{lyndenbell62}. Alternatively, the 
disk may exhibit large amplitude bursts of accretion, or it may 
fragment and break up into distinct bound objects. The first 
two modes are relevant to disk evolution, 
while the possibility of fragmentation is of great interest 
as a mechanism for forming massive planets or 
substellar objects \citep{boss97,stamatellos07,boley09}.

Although the local linear stability of a gas disk is only a function of its 
instantaneous structure -- the surface density and sound speed -- 
the non-linear evolution 
depends critically upon the thermodynamic and radiative properties 
of the disk \citep{pickett98,pickett00,nelson00}. The importance of cooling 
for the non-linear behavior can be demonstrated, in part, using elementary 
arguments, provided that we stipulate that self-gravitating disk turbulence acts 
as a local process \citep{paczynski78}. In this limit, the constraint 
implied by the linear stability threshold (Equation~\ref{eq:2_toomre}) 
allows us to 
determine an explicit form for $\alpha(\Sigma,\Omega_{\rm K})$. We first 
note that for a disk primarily heated by viscous dissipation, rather than 
irradiation, thermal equilibrium implies that the heating rate per unit area, 
$(9/4) \nu \Sigma \Omega_{\rm K}^2$, must balance cooling at a rate $2 \sigma T_{\rm eff}^4$. 
In terms of $\alpha$,
\begin{equation}
 \alpha = \frac{4}{9 \gamma (\gamma-1)} \frac{1}{t_{\rm cool} \Omega_{\rm K}},
\label{eq:3_alpha_tcool} 
\end{equation}
where $\gamma$ is the adiabatic index, $t_{\rm cool}  = U / 2 \sigma T_{\rm eff}^4$ and $U$, the 
thermal energy per unit surface area, is given by $U = c_s^2 \Sigma / 
\gamma (\gamma-1)$. This form for $\alpha(t_{\rm cool})$ is an  
identity for viscous disks \citep[see e.g.][ \S7]{pringle81}, but it 
does not generally determine $\alpha$ because $T_{\rm eff} (\Sigma)$, 
which is required to determine $t_{\rm cool}$, is unknown. For a 
disk in which self-gravity is the sole source of angular momentum 
transport, however, we have the advantage of an additional constraint; 
$\Sigma$ must be related to the central 
sound speed via Equation~(\ref{eq:2_toomre}). If we take 
$Q_0$, the value of the Toomre 
parameter in the saturated state of gravitational instability,  
to be a constant (thereby anticipating the result that a self-gravitating 
disk tends to maintain a state of marginal instability with $Q_0 > 1$), 
then $c_s = \pi G Q_0 \Sigma / \Omega_{\rm K}$. 
Making this assumption, we use the 
one-zone approximation to the equation of radiative diffusion 
in $z$ to relate the central temperature, $T_c$, to the effective 
temperature $T_{\rm eff}$. For an optically thick disk \citep[for details, see 
e.g.][]{hubeny90},
\begin{equation}
 T_c^4 \sim \tau T_{\rm eff}^4,
\end{equation}
where $\tau$, the optical depth between the disk surface and 
mid-plane, is given in terms of the mid-plane opacity by 
$\tau = (1/2) \Sigma \kappa$. Writing $\kappa = \kappa_0 T_c^\beta$, 
one finds that $\alpha$ can be expressed as an explicit function 
of just the surface density and angular velocity \citep{levin07},
\begin{equation} 
 \alpha \sim \frac{\sigma}{\kappa_0} 
 \left( \frac{\mu m_p}{\gamma k_B} \right)^{4-\beta}
 \left( Q_0 \pi G \right)^{6 - 2 \beta} 
 \Sigma^{4-2 \beta} \Omega^{2 \beta - 7}.
\label{eq:3_alpha_sg} 
\end{equation} 
Here we have omitted numerical prefactors, the exact values of 
which vary in the literature depending upon the assumed vertical 
structure (which has not been accurately determined from 
simulations), and  with whether the sound 
speed used is isothermal or adiabatic. Analogous expressions can 
be derived for optically thin disks, or for 
disks with more general opacity laws that are a function of density 
as well as temperature \citep{cossins10}.

The simplicity of Equation~\ref{eq:3_alpha_sg} results from the 
assumption that angular momentum transport due to self-gravity 
can be treated as a local, ``gravitoturbulent", process. Since 
gravity is a long-range force, the general validity of the local 
approximation is questionable, and it is easy to envisage circumstances 
in which it would fail entirely. A disk that develops a strong 
bar-like structure, for example, could have regions in which 
gravitational torques from the bar remove angular momentum 
and drive accretion without the attendant energy dissipation 
that occurs in a local model. More formally, \citet{balbus99} 
show that a self-gravitating disk can only be described by a 
local $\alpha$ model if, at all radii, the pattern speed of the 
waves $\Omega_p$ matches the local angular velocity 
$\Omega$. If, on the other hand, $\Omega_p$ is very 
different from $\Omega$, the radial energy flux associated 
with waves in the disk will violate the assumption of locality and 
mandate a global treatment of the disk dynamics.

The locality of angular momentum transport by disk self-gravity 
has been tested numerically. \citet{lodato04,lodato05} studied 
angular momentum transport in a model system composed of 
an isolated disk that cools everywhere on a time scale that is a 
fixed multiple of the local dynamical time. They found that the 
analytic estimate for the stress, 
derived from the assumption of local thermal balance, was a 
good match to the numerical results provided that the disk 
mass was not too high, $M_{\rm disk} / M_* \lesssim 0.25$. 
A more quantitative analysis was carried out by \citet{cossins09},  
who explicitly evaluated the pattern speed of the waves generated by 
self-gravity within a disk described using a similarly simple 
prescription for the cooling time. For $M_{\rm disk} / M_* = 0.1$, 
they bounded departures from local behavior at the 10\% level. 
Broadly similar results also hold for more realistic disks, in 
which the cooling time scale, set by the radiative physics, 
is a strong function of radius. \citet{boley06} used a radiation 
hydrodynamics code to study the evolution of a moderately 
massive disk ($M_{\rm disk} / M_* = 0.14$) around a low 
mass star. They found that, although some non-local transport 
did occur, the predictions of a local model based upon the 
cooling time were consistent with the simulation across a 
substantial region of the disk. Likewise, simulations including 
radiative transfer by \citet{forgan10} showed significant non-local 
effects only when $M_{\rm disk} / M_* \gtrsim 0.5$.

The conditions for fragmentation of a low-mass self-gravitating disk 
can also be written, to a first approximation, as a function of local disk 
conditions. \citet{gammie01}, using local numerical simulations, 
found that if the cooling could be parameterized as,
\begin{equation}
 t_{\rm cool} = \beta \Omega_{\rm K}^{-1},
\end{equation}
then for $\beta \lesssim 3$ fragmentation followed promptly as 
soon as the disk cooled to low $Q$. Short 
cooling times allow the disk to radiate away the energy provided 
by shocks or turbulent dissipation, such that neither pressure support nor 
shear suffice to avert gravitational collapse \citep{shlosman87}. 
The thermal condition for fragmentation in this simple form applies 
strictly only in the limit where the cooling time in the disk is locally 
constant. Using a more realistic model, in which the cooling was 
calculated using a one-zone vertical model that took account of the 
optical depth, \citet{johnson03} found that strong variations in the opacity 
with temperature could result in large changes to the critical cooling 
time scale for fragmentation. In practice, however, the opacity in 
protoplanetary disks is a (relatively) slowly varying function for the 
temperatures, below the dust sublimation temperature $T \approx 
1500~{\rm K}$, that are most likely to coincide with self-gravitating 
regions. Large deviations from the predictions of constant cooling 
time models are thus only likely for very high accretion rates 
$\dot{M} \gtrsim 10^{-5} \ M_\odot \ {\rm yr}^{-1}$ \citep{cossins10}.

It is unclear how accurately the local thermal criterion for fragmentation -- derived from 
two dimensional simulations -- holds in three dimensions in global disks. 
Early global simulations found surprisingly good quantitative agreement. 
\citet{rice03}, using Smooth Particle Hydrodynamics (SPH) simulations, found that 
for low-mass disks (roughly those for which $M_{\rm disk} / M_* \sim 0.1$, 
or less), the condition to avoid fragmentation could indeed be expressed via a 
minimum cooling time, given for $\gamma = 5/3$ by $\beta \approx 5$. 
Equivalently (if irradiation is not important) disks can stably transport 
angular momentum in a self-regulated state without fragmenting 
up to a maximum $\alpha \simeq 0.1$ 
\citep{rice05}. Similarly, \citet{mejia05}, using a grid-based hydrodynamics 
code, did not obtain fragmentation for a cooling time $\beta \approx 6$. 
Recent higher resolution SPH simulations, however, see evidence for 
fragmentation at significantly longer (by a factor of two) cooling times 
\citep{meru10}. Additional work, ideally using independent numerical 
techniques, is needed 
to clarify the relation between the two dimensional local and fully 
three dimensional fragmentation boundary.

\begin{figure}[t!]
\includegraphics[width=\textwidth]{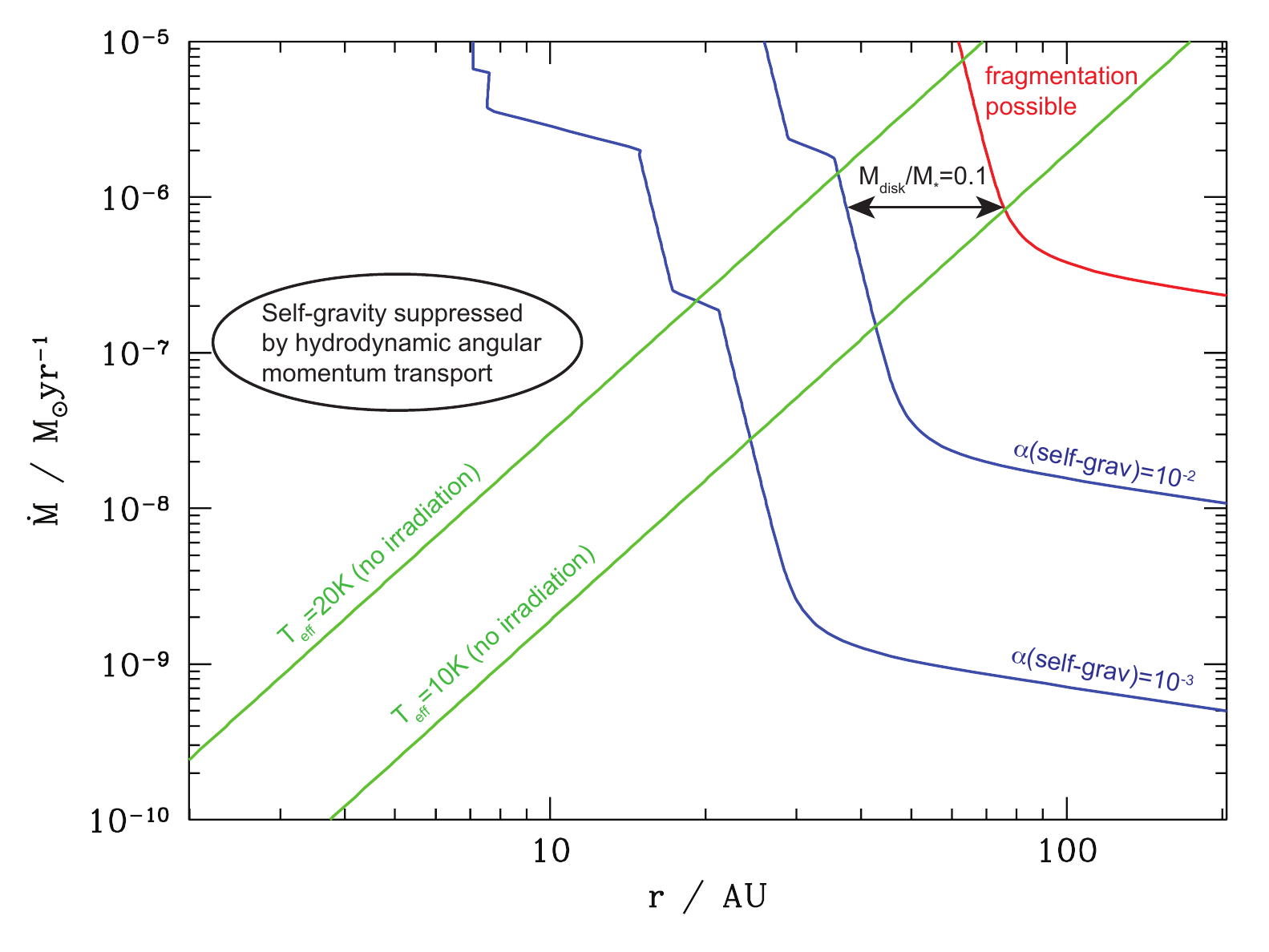}
\caption{The structure of a steady-state self-gravitating disk around a 
$1 \ M_\odot$ star, calculated assuming that self-gravity acts as a local 
angular momentum transport process whose efficiency is set by the 
requirement of thermal equilibrium \citep{clarke09,rafikov09}. The opacity 
includes contributions from water ice, amorphous carbon, silicates and graphite (Z. Zhu, 
private communication).  If self-gravity is the sole source of 
angular momentum transport, the effective $\alpha_{\rm grav}$ generated by the self-gravitating 
turbulence is an increasing function of radius, and there is a maximum radius 
(shown as the red line) beyond which fragmentation is predicted to occur. The blue 
curves show radii where $\alpha_{\rm grav}=10^{-2}$ and $10^{-3}$. If 
other sources of angular momentum transport co-exist with self-gravity, and generate 
stresses of this magnitude, these radii denote the approximate inner edge of the 
self-gravitating disk. The green lines show where the effective temperature of 
the luminous disk equals 10~K or 20~K; if an external irradiation field 
heats the disk it would affect the dynamics to the right of the 
green lines.}
\label{fig:3_selfgrav}
\end{figure}

The numerical results imply that for a low 
mass disk both the angular momentum transport efficiency of a stable 
self-gravitating disk, and the conditions for fragmentation, can be written 
as functions of the angular velocity and local opacity. It is then possible 
to abstract the physics of disk self-gravity into a one-dimensional 
model \citep[identical in spirit to that of][]{lin87}, in which $\nu(r,\Sigma)$ is completely 
determined by the requirement of thermal equilibrium, and seek either 
steady-state analytic \citep{matzner05,clarke09,rafikov09,cossins10} or 
time-dependent numerical solutions \citep{rice09,zhu10}. A central result of such simplified models is 
that fragmentation cannot occur in the inner disk \citep{rafikov05}. 
\citet{matzner05}, for example, derived a critical angular velocity for 
fragmentation of $\Omega_{\rm frag} \simeq 
4.3 \times 10^{-10} \ {\rm s}^{-1}$, corresponding to a radius of 60~AU 
around a Solar-mass star. Results from a similar calculation are plotted 
in Figure~\ref{fig:3_selfgrav}, which shows the region in the 
$(r,\dot{M})$ plane for which a non-irradiated self-gravitating disk 
would be unable to maintain self-regulated transport, and might 
fragment. In agreement with numerous 
authors \citep{clarke09,rafikov09,zhu10,cossins10}, we 
find that disks with $\dot{M} \gtrsim 10^{-7} \ M_\odot 
\ {\rm yr}^{-1}$ can be unstable to fragmentation (or, some other 
breakdown of local behavior, such as bursts) at radii beyond 
50 -- 10$^2$~AU, but are stable closer in. This result is consistent 
with several simulations of protoplanetary disks 
that include radiative transfer \citep[e.g.][]{boley10,stamatellos08}, as expected given 
that the assumptions underlying the analytic theory were derived 
numerically. Some other radiation hydrodynamic simulations, however, 
obtain fragmentation at significantly smaller radii $r \sim 10 \ {\rm AU}$ 
\citep{boss08,mayer07}. The reason for these discrepant results is 
not fully understood, although differences in the treatment of 
the chemical composition (e.g. the temperature dependence of $\mu$) 
and radiative transfer may play a role \citep{cai10}.

In the context of star formation, 50 -- 10$^2$~AU 
is not a particularly large radius, and the analytic result then suggests 
molecular cloud cores with sufficiently high specific angular momentum 
would be susceptible to fragmentation at their outer edge 
\citep{matzner05,stamatellos07,stamatellos09,rice10,kratter10b}. There are, however, two caveats. First, 
the effective temperature of an isolated self-gravitating disk at the 
innermost fragmenting radius is quite low, typically around 
10-20~K \citep{matzner05}. This implies that the dynamics of 
self-gravitating disks in the potentially fragmenting region are 
sensitive to modest levels of external irradiation, which will 
tend to weaken gravitational instabilities \citep{cai08,stamatellos08,vorobyov10}. 
Second, the innermost fragmenting radius is calculated for a 
given opacity, and will vary with the metallicity of the disk and 
as a result of particle coagulation \citep{cai06}. For an opacity appropriate 
for ice grains, for example, $\kappa = \kappa_0 T^2$ \citep{semenov03}, 
and $\Omega_{\rm frag} \propto \kappa_0^{-1/3}$ \citep{matzner05,rafikov09}. 
Reduced values of the opacity would therefore push the fragmentation 
boundary interior to the fiducial radius of 50 -- 10$^2$~AU.

Immediately interior to the radius where $\Omega_{\rm K} = \Omega_{\rm frag}$, 
the strength of self-gravitating transport is close to the maximum value, 
$\alpha \approx 0.1$. At smaller radii the disk becomes increasingly 
optically thick, and the longer cooling time results in lower values of 
$\alpha$ (Equation~\ref{eq:3_alpha_tcool}). This means that the surface 
density in self-gravitating disks at small radii is high, and that even very 
weak transport by other mechanisms will suffice to 
render the disk stable against self-gravity. Curves corresponding to 
a self-gravitating $\alpha = 10^{-2}$ and $10^{-3}$ are shown in 
Figure~\ref{fig:3_selfgrav}, from which it is clear that is hard to 
sustain a strongly self-gravitating disk much interior to 10~AU. 
This implies that, although shocks in a strongly self-gravitating disk 
are a candidate mechanism for chondrule formation \citep{boss05}, 
it is hard to construct self-consistent models in which strong enough 
shocks would be present at small radii \citep{boley08}.

\begin{figure}[t!]
\includegraphics[width=\textwidth]{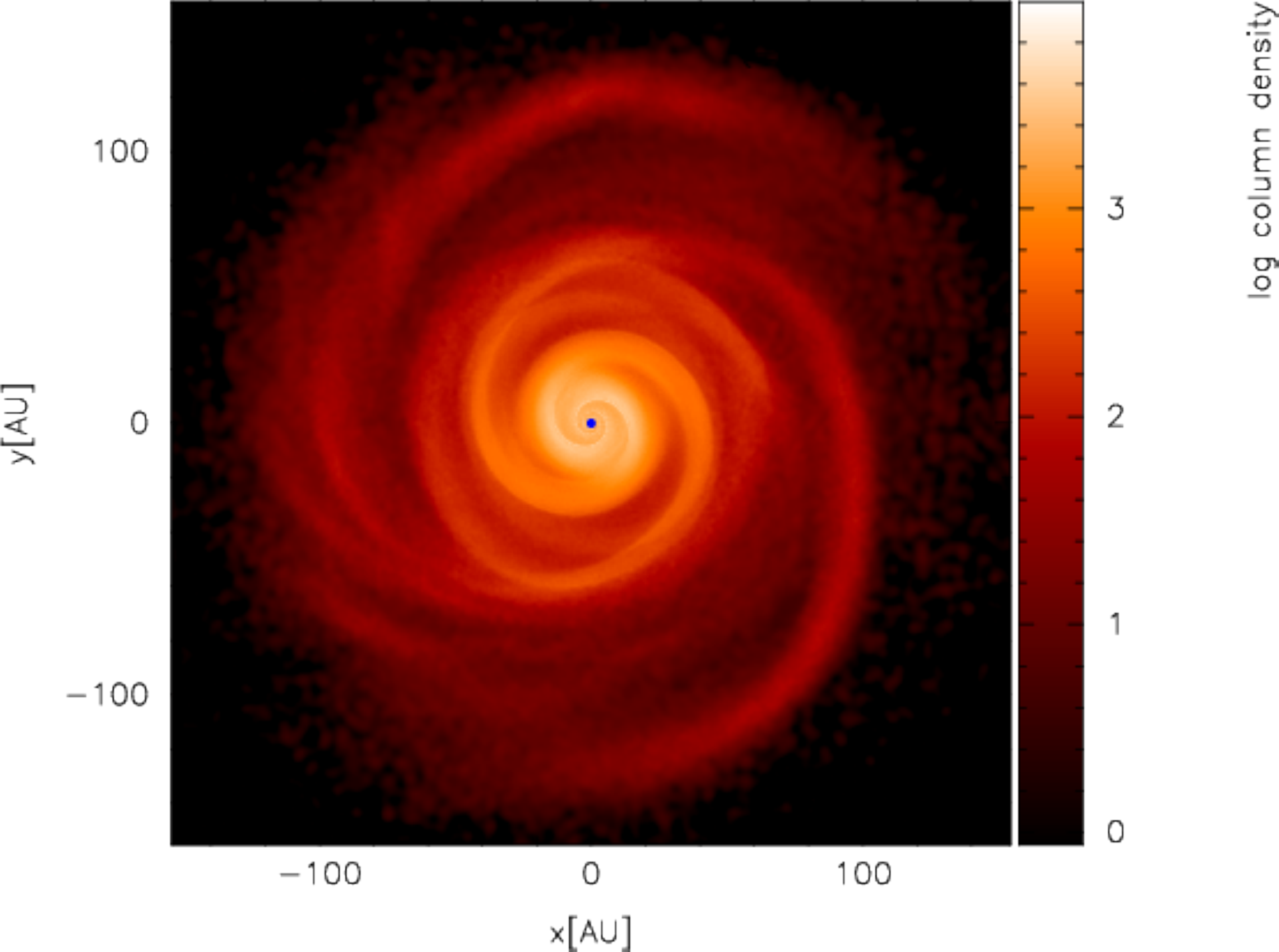}
\caption{Structure of a massive ($M_{\rm disk} = M_* = M_\odot$) self-gravitating disk 
in the non-fragmenting regime, based on simulations by \citet{forgan10}. The 
simulated disk had an initial surface density profile $\Sigma \propto r^{-3/2}$, 
and was evolved with an approximate radiative transfer scheme. At 
this mass ratio, strong low-order spiral structure dominates.}
\label{fig:3_spirals}
\end{figure}

For more massive isolated disks, the nature of the departure from 
local behavior is reasonably well-established. As the mass ratio 
between the disk and star increases, there is an increasing 
domination of low-order global spiral modes (illustrated in 
Figure~\ref{fig:3_spirals}), and a transition to  
time-dependent ``bursting" behavior \citep{vorobyov06}. For protostellar 
disks at early times, however, multiple confounding factors come into play, 
including irradiation from the envelope, and dynamical effects 
associated with the addition of mass \citep{harsono10}, angular momentum, and 
thermal energy from infall. All of these will affect the self-gravitating evolution.

\citet{kratter10} simulated the evolution of protostellar disks 
formed from rapid infall, assuming that the disk  
behaves isothermally prior to fragmentation. The 
assumption of isothermality precludes comparison with the 
results for low mass, isolated disks -- where thermal physics is 
of primary importance -- though it represents a reasonable starting point 
for the background structure of disks whose thermal balance is dominated by 
external irradiation 
\citep{krumholz07}. \citet{kratter10} express the rate of infall via a parameter,
\begin{equation}
 \xi = \frac{\dot{M}_{\rm in} G}{c_s^3},
\end{equation} 
where $\dot{M}_{\rm in}$ is the rate of infall onto the disk and $c_s$ 
is the disk sound speed. Since the rate of accretion through a 
self-gravitating disk can be written as $\dot{M} = 3 \alpha c_s^3 / (GQ_0)$, 
the parameter $\xi$ represents the infall rate scaled to the maximum 
(at $\alpha \sim 1$) accretion rate that a self-gravitating disk might be 
expected to sustain. Numerically, it is found that 
disks are able to avoid fragmentation for $\xi \lesssim \xi_{\rm max}$, 
where $\xi_{\rm max}$ varies with the angular momentum of the 
infalling gas but is typically $\xi_{\rm max} \sim 2-3$ \citep{kratter10}. The 
corresponding $\alpha$, averaged over the disk, is $\alpha \approx 1$, 
significantly higher than the maximum value for an isolated low-mass 
disk. Accretion in this regime is highly variable, 
due to a combination of the dominance of global modes \citep{laughlin94,forgan10}, 
the effect of infall, and the formation and disruption of clumps 
\citep{vorobyov10,boley10}. Disks can attain masses as high as 
$M_{\rm disk} \approx M_*$ while remaining in the stable regime; 
yet more massive disks invariably fragment.

\subsection{Magnetorotational Instability}
\label{sec:3_intro}
An accretion disk coupled dynamically to a weak magnetic field is 
subject to instabilities that initiate and sustain MHD turbulence, which 
in turn generates Maxwell and Reynolds stresses that result in outward 
angular momentum transport. In ideal MHD (i.e. when terms associated 
with Ohmic and ambipolar diffusion, and with the Hall effect, can be 
neglected), the conditions necessary for this instability -- known as the 
magnetorotational instability \citep[MRI;][]{balbus91} -- are simply stated 
\citep[for an excellent review, see][]{balbus98}. A disk threaded by a vertical 
magnetic field, that is ``weak" in the sense that,
\begin{equation}
 \frac{B_z^2}{8 \pi} \lesssim \frac{3}{\pi^2} \rho c_s^2,
\end{equation}
is subject to a local, linear instability, provided that,
\begin{equation}
 \frac{\rm d}{{\rm d}r} \left( \Omega^2 \right) < 0.
\end{equation} 
Keplerian disks are therefore always unstable to the MRI, unless they 
are threaded by strong magnetic fields whose energy density exceeds 
the thermal energy within the disk \citep[even then, the MRI could 
arise as a secondary instability if the strong field is unstable to, say, 
Parker instability;][]{johansen08}. Although there is no observational 
proof, the ubiquity of the conditions necessary for the MRI, together 
with its fast growth rate ($\sim \Omega$ for the fastest growing mode), 
strongly suggest that the MRI is the source of angular momentum transport in 
non-self-gravitating disks around white dwarfs, neutron stars, and black holes.

For protoplanetary disks, no comparably definitive statement is possible. 
The gas in protoplanetary disks is cool, dense, and overwhelmingly neutral, 
and as a result non-ideal MHD effects are important. Complicating matters 
further, different non-ideal terms dominate at different $(r,z)$ within the 
disk. The strength and nature of turbulence 
sustained by the MRI is then not expected to be universal, but rather must 
be calculated as a function of the local physical conditions. (It is also 
possible that some disks might be threaded by magnetic fields that are 
too strong to admit the MRI altogether.) 

Although the full problem is complex, considerable insight can be gained 
by examining a toy model in which the sole non-ideal effect is Ohmic resistivity 
\citep[the discussion here is adapted from][]{gammie96}. The induction equation 
has the form,
\begin{equation}
 \frac{\partial {\bf B}}{\partial t} = 
 \nabla \times \left[
  {\bf v} \times {\bf B} - \eta \nabla \times {\bf B} \right],
\end{equation}
where $\eta$, the magnetic resistivity, enters as a diffusive term that acts 
to smooth out small scale structure in the magnetic field. To estimate 
the effect of resistivity on the MRI, we note that the time scale for the 
growth of the MRI on a scale $\lambda$ in ideal MHD is 
$\tau_{\rm MRI} \sim \lambda / v_A$, where $v_A$, the Alfv\'en 
velocity, is given by $v_A = B / \sqrt{4 \pi \rho}$. At the same scale, 
the time scale for resistive damping is $\tau_\eta \sim \lambda^2 / \eta$. 
Equating these time scales at $\lambda = h$ (which is where the effects 
of non-zero $\eta$ are weakest), we find that to suppress the growth of the 
MRI requires,
\begin{equation}
 \eta \gtrsim h v_A.
\end{equation}
Defining the magnetic Reynolds number, ${\rm Re}_M = h v_A / \eta$, 
an equivalent statement is that the MRI will not exist for ${\rm Re}_M \lesssim 1$ 
(note that care is needed when reading the literature, since at least three related but distinct definitions of ${\rm Re}_M$ are in common currency). 

Next, we estimate the electron fraction needed to attain ${\rm Re}_M = 1$ under 
disk conditions. In a gas composed of electrons and neutral particles, collisions 
between the two species limit the conductivity and yield a resistivity that is 
given by \citep{spitzer62,blaes94},
\begin{equation}
 \eta = 234 \left( \frac{n}{n_e} \right) T^{1/2} \ {\rm cm}^2 \ {\rm s}^{-1},
\label{eq:3_eta} 
\end{equation}
where $n$ and $n_e$ are the number densities of the neutral species and 
electrons, respectively. For a magnetic field strength such that $v_A = \epsilon c_s$, 
the critical electron fraction necessary to reach ${\rm Re}_M = 1$ at 1~AU can 
be estimated as,
\begin{equation}
 \frac{n_e}{n} \sim 5 \times 10^{-13} \left( \frac{h/r}{0.05} \right)^{-1} 
 \left( \frac{\epsilon}{0.1} \right)^{-1}.
\label{eq:3_xcrit} 
\end{equation} 
This is a very small number. It is, however, much larger than the ionization 
fraction for disk gas in thermal equilibrium at a few hundred K. Apart from 
very close to the star (where thermal ionization of alkali metals provides 
sufficient ionization), the operation of the MRI in protoplanetary disks 
therefore depends on non-thermal ionization processes. For a particularly 
simple example, we can evaluate the electron fraction expected if the 
dominant source of ionization is the decay of $^{26}{\rm Al}$. To an order 
of magnitude, the expected ionization rate is $\zeta = 10^{-19} \ {\rm s}^{-1}$ 
\citep{stepinski92}. Balancing this against dissociative recombination in the 
gas phase, we find,
\begin{equation}
 \frac{n_e}{n} = \sqrt{ \frac{\zeta}{\beta n} } 
 \simeq 10^{-13} \left( \frac{\zeta}{10^{-19} \ {\rm s}^{-1}} \right)^{1/2} 
 \left( \frac{T}{500 \ {\rm K}} \right)^{1/4}
 \left( \frac{n}{10^{14} \ {\rm cm}^{3}} \right)^{-1/2}, 
\label{eq:3_xactual} 
\end{equation}
where we have assumed a rate coefficient 
$\beta = 3 \times 10^{-6} T^{-1/2} \ {\rm cm}^3 \ {\rm s}^{-1}$ 
\citep[e.g.][]{fromang02}, and substituted parameters 
appropriate for the disk mid-plane near 1~AU.

Comparing Equations~(\ref{eq:3_xcrit}) and (\ref{eq:3_xactual}) shows 
why determining the efficacy of the MRI in protoplanetary disks is a 
hard problem; the electron fraction that is expected to be present due 
to non-thermal ionization can be of the same order of magnitude as 
the simple estimate of the critical fraction needed to sustain the MRI. 
Obtaining the right answer requires detailed calculations that address 
each of the places where we resorted to order of magnitude arguments above:
\begin{enumerate}
\item
What is the effect of non-ideal terms on the MRI? Specifically, how 
do Ohmic and ambipolar diffusion, and the Hall effect, alter the linear and 
non-linear development of MRI-driven disk turbulence?
\item
What is the spatial distribution of ionization within the disk, due to non-thermal 
processes including X-ray irradiation, cosmic rays, and radioactive decay?
\item
What is the recombination rate, due to both gas phase and surface processes 
(on dust grains)?
\end{enumerate}
Despite substantial progress on each of these individual issues, discussed 
below, the compounded uncertainty involved when they are combined 
precludes a definitive assessment of how efficient the MRI is at transporting 
angular momentum in protoplanetary disks. The implicit coupling to the 
physics of dust coagulation, that enters via the influence of dust on the 
recombination rate, is particularly troubling, as dust coagulation is subject 
to numerous additional uncertainties.

\subsubsection{Effects of non-ideal terms}
The evolution of the magnetic field within a weakly ionized protoplanetary 
disk obeys an induction equation that reads,
\begin{equation}
 \frac{\partial {\bf B}}{\partial t} = \nabla \times 
 \left[ {\bf v} \times {\bf B} 
 - \eta \nabla \times {\bf B} 
 - \frac{ {\bf J} \times {\bf B} }{e n_e} 
 + \frac{ ( {\bf J} \times {\bf B} ) \times {\bf B} }{c \gamma \rho_i \rho} \right].
\end{equation} 
Here ${\bf v}$ is the velocity of the dominant neutral component of the 
fluid, with density $\rho$, the current ${\bf J} = c (\nabla \times {\bf B}) / 4 \pi$, 
and $\rho_i$ is the density of ions. $\gamma$, the drag coefficient, is 
given in terms of the ion-neutral collision rate $\langle \sigma \omega \rangle_i$ 
by $\gamma = \langle \sigma \omega \rangle_i / (m_i + m_n)$, where $m_i$ and 
$m_n$ are the masses of the ions and neutrals respectively \citep[for a derivation see, 
e.g.][]{balbus10}. The terms on the right-hand-side describe 
magnetic induction (the frozen-in field behavior of ideal MHD) and the three 
non-ideal effects, Ohmic diffusion (denoted as O), the Hall effect (H), and 
ambipolar diffusion (A). Physically, ambipolar diffusion is dominant 
when the field is well-coupled to the ions and electrons, such that the field 
drifts with the charged species relative to the neutral component. Ohmic 
diffusion dominates when the conductivity is so low that the field is 
imperfectly coupled to both the electrons and the ions. Finally, the Hall 
effect is most important in an intermediate regime where the field is 
well-coupled to the electrons but not to the ions.

\begin{figure}[t!]
\includegraphics[width=\textwidth]{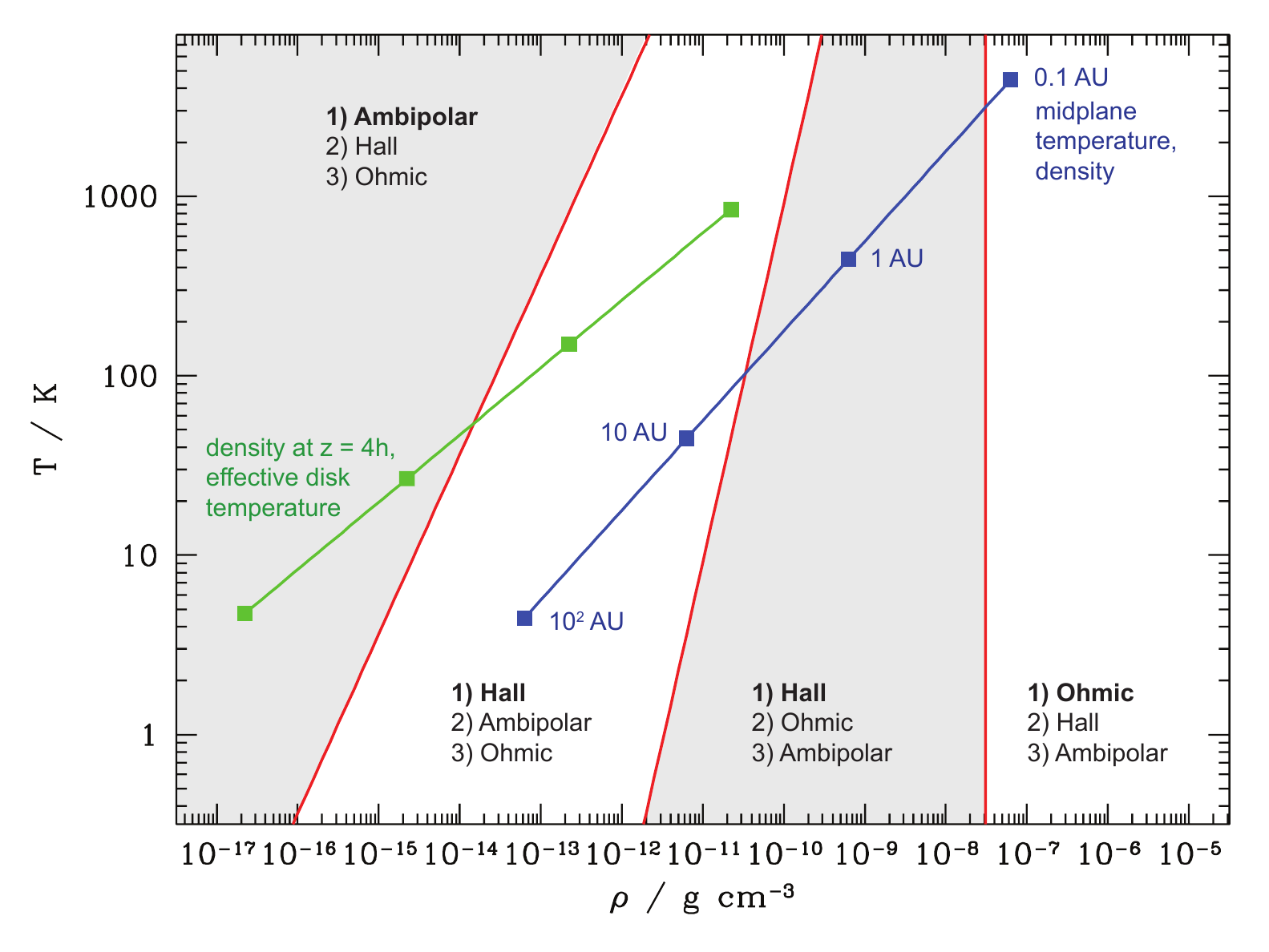}
\caption{The relative importance of non-ideal MHD terms is shown in the 
$(\rho,T)$ plane \citep{balbus01,kunz04}, assuming a magnetic field 
strength such that $v_A / c_s = 0.1$. Also plotted are very approximate tracks 
showing the radial variation of physical conditions at the mid-plane, and near 
the surface, of protoplanetary disks. The mid-plane conditions are estimated 
for a disk around a Solar mass star with $\Sigma = 10^3 (r / 1~{\rm AU})^{-1} \ 
{\rm g \ cm}^{-2}$, 
and $(h/r) = 0.04$. The surface conditions are estimated from the density 
at $z = \pm 4 h$ (using a gaussian density profile), assuming that the 
temperature is the effective temperature for a steady-state disk accreting 
at $\dot{M} = 10^{-7} \ M_\odot {\rm yr}^{-1}$.}
\label{fig:3_nonideal}
\end{figure}

Determining the absolute importance of the non-ideal terms (i.e. their 
ratios to the inductive term) requires solving for the ionization state 
of the disk. As we have already observed, this is difficult everywhere 
except in the very innermost regions, interior to about 0.1~AU, where 
thermal ionization dominates. It is much easier to assess the relative 
magnitude of the non-ideal terms, which depend only upon the 
temperature, $T$, and total number density $n$. \citet{balbus01} 
estimate these ratios by assuming that electrons and singly-ionized 
ions are the charge carriers, that the typical fluid velocities are 
$\sim v_A$, the Alfv\'en speed, and that typical gradients are $\sim h^{-1}$. 
They obtain,
\begin{eqnarray} 
\frac{\rm O}{\rm H} & = & 
 \left( \frac{n}{8 \times 10^{17} \ {\rm cm}^{-3} } \right)^{1/2} 
 \left( \frac{v_A}{c_s} \right)^{-1} \\
 \frac{\rm A}{\rm H} & = & 
 \left( \frac{n}{9 \times 10^{12} \ {\rm cm}^{-3} } \right)^{-1/2} 
 \left( \frac{T}{10^3 \ {\rm K}} \right)^{1/2} 
 \left( \frac{v_A}{c_s} \right).
\end{eqnarray} 
Using these expressions, we show in Figure~\ref{fig:3_nonideal} the 
relative importance of the three non-ideal effects as a function of 
density and temperature \citep[after][]{kunz04}. Over-plotted on the 
Figure are order of magnitude estimates of the run of density and 
temperature within a disk, both at the mid-plane and near the 
surface. At the disk mid-plane, 
a combination of the Hall effect and Ohmic diffusion are likely 
to be the most important non-ideal terms interior to $\sim 10$~AU. 
At larger mid-plane radii -- and in the disk atmosphere at all radii -- 
the Hall effect and ambipolar diffusion are dominant.

The linear stability of disks to growth of the MRI has been studied 
for each of the relevant combinations of non-ideal terms. The simplest 
case is the purely resistive limit, analyzed by \citet{jin96}, whose 
mathematical behavior closely matches that expected from the 
heuristic arguments given above. If Ohmic diffusion is the sole 
non-ideal effect, the growth of the MRI is damped whenever 
${\rm Re}_M \lesssim 1$ \citep{jin96,sano99}. More typically, 
however, Ohmic diffusion will be accompanied by a strong 
Hall effect, which qualitatively modifies the disk's 
linear stability \citep{wardle99,balbus01}. Most 
strikingly, for a Keplerian disk in which $\kappa = \Omega_{\rm K}$, if
\begin{equation}
 {\rm Re}_M^{-1} < \frac{3}{4} 
 \left( \frac{v_A}{c_s} \right),
\end{equation} 
i.e. for disks in which Ohmic diffusivity is not extremely 
strong, the Hall effect destabilizes some perturbations on 
arbitrarily small scales even in the presence of non-zero $\eta$. 
Other perturbations are stabilized. In general, for a perturbation 
with wavevector $\bf k$, the Hall term acts to stabilize or destabilize 
depending upon the sign of $({\bf k} \cdot {\bf \Omega})({\bf k} \cdot {\bf B})$ 
\citep{balbus01}. If vertical fields thread the disk, one may conjecture 
that the resulting asymmetry in the dynamics -- between fields that are aligned 
or anti-aligned with the angular momentum vector -- could 
have physical consequences.

In addition to the Hall / Ohmic-dominated regime, the case where ambipolar 
diffusion is the strongest non-ideal effect may also be relevant to 
protoplanetary disks. The linear physics of the MRI in this limit has 
been analyzed using a two-fluid model by \citet{blaes94}, and in a 
single-fluid approximation by \citet{kunz04} and by \citet{desch04}. 
The novel feature of the ambipolar MRI is that -- even in the absence of 
the Hall effect -- the addition of what is an apparently dissipative term can 
destabilize fluid configurations that would be stable in purely ideal 
MHD \citep{kunz04}. Nevertheless, it remains true that the growth rate 
of the ambipolar MRI is strongly suppressed once the ion-neutral collision 
time scale exceeds $\Omega_{\rm K}^{-1}$. As with Ohmic diffusion, 
one can therefore define an ``ambipolar Reynolds number", 
${\rm Re}_A = \gamma \rho_i / \Omega_{\rm K}$, and express the 
condition for efficient MRI growth simply as requiring 
${\rm Re}_A \gtrsim 1$.

\subsubsection{Non-linear MRI evolution}

The linear physics of the MRI, in the various non-ideal regimes, plausibly 
defines a set of necessary conditions for the onset of MHD turbulence and 
angular momentum transport. It is less clear whether linear instability 
generally defines sufficient conditions for sustained transport, since, 
for a disk with zero net magnetic flux, no conservation law prohibits 
the disk from reconnecting its initial field and thereby accessing a 
hydrodynamically stable, field-free state. \citep[This curious behavior 
is seen in MHD simulations in small domains that omit explicit dissipation, e.g.][]
{pessah07,fromang07,guan09,simon09b}. Numerical simulations to study 
the non-linear phase  
can be carried out in either a local (``shearing box") geometry, which 
amounts to modeling a small patch of disk in a co-rotating co-ordinate 
system \citep{hawley95}, or globally \citep{armitage98}. A secondary 
distinction is between unstratified simulations, which by omitting the vertical 
component of gravity have no mean vertical density gradient (this is 
done for reasons of computational economy), and stratified calculations 
\citep{brandenburg95,stone96} that are required in order to model physical 
effects such as buoyancy.

A tentative understanding 
of the non-linear behavior of the MRI has been attained only in the ideal MHD limit. 
In this regime, stratified local simulations confirm that the MRI yields 
sustained MHD turbulence, in which outward angular momentum transport 
occurs primarily as a result of Maxwell stresses (though fluid stresses also 
contribute at a non-negligible level). The toroidal magnetic field is dominant, 
and displays quasi-periodic cyclic behavior near the mid-plane on a 
time scale of $\sim 10$ orbits \citep{brandenburg95,stone96,davis10,shi10}. 
Associated with this cycle, magnetic field rises buoyantly through the 
disk, forming a ``butterfly diagram" in two-dimensional plots of 
$B_\phi (z,t)$ that is strikingly reminiscent of Solar dynamo cycles. 
Typical values of $\alpha$ in zero net field simulations, averaged 
over long time scales, are $\alpha \sim 10^{-2}$ \citep{davis10,shi10}. 
Larger values are possible  if a net field threads the disk, 
in which case unstratified simulations show that the 
stress scales with $\beta$, the ratio of gas pressure to 
the magnetic pressure associated with the net flux, 
as $\alpha \propto \beta^{-1/2}$ \citep{hawley95}. Some stratified simulations 
suggest that a net field may also 
promote mass loss via a disk wind \citep{suzuki09}.

Although the basic results summarized above appear robust, loose 
ends of unknown physical significance are present even in 
``almost ideal" MHD, where viscosity and resistivity are 
introduced solely to give well-controlled dissipation at the grid scale.
In particular, the saturation level (and even 
the persistence) of turbulence has been observed to depend upon the magnetic 
Prandtl number ${\rm Pm} = \nu / \eta$, even at relatively high Reynolds 
numbers where Ohmic damping of linear modes should not be playing a 
role \citep{fromang07b,longaretti10,simon09}. Whether this dependence 
persists at the Reynolds numbers of real disks (which are vastly larger 
than those accessible computationally) remains an open question.

Moving beyond ideal MHD, exploratory local simulations have studied 
the non-ideal regimes relevant to protoplanetary disks. Caution 
is required in interpreting many of the results, since 
the bulk of the work predates the realization that unstratified local 
simulations with zero-net flux exhibit pathological convergence properties. 
Nonetheless, simulations that include Ohmic diffusion confirm the 
analytic expectation of a critical Reynolds number 
below which MRI activity is quenched \citep{fleming00,sano01}, and 
suggest that the primary criterion for quenching continues to be set by the 
resistivity even in the presence of the Hall effect \citep{sano02}. \citet{simon09}, 
for example, quantified the conditions necessary for MRI activity in 
unstratified local simulations with a net toroidal field, including both 
resistivity and an explicit microphysical viscosity. They found 
that turbulence decayed when a dimensionless measure of the importance of 
resistivity, the Elsasser number,
\begin{equation}
 \Lambda \equiv \frac{v_A^2}{\eta \Omega_{\rm K}} \lesssim 10^2.
\end{equation} 
For somewhat weaker resistivity, additional time-dependent behavior, 
over and above that seen in ideal MHD, develops \citep{simon10}. 
A cycle develops in which the mid-plane MRI is first quenched, but 
subsequently reignites due to long time scale growth of the toroidal 
magnetic field. 
Lastly, in the ambipolar regime, numerical work by \citet{maclow95} and by 
\citet{hawley98} shows that the MRI is suppressed for 
ion-neutral collision frequencies below about $10^2 \ \Omega_{\rm K}$.

By any accounting our quantitative understanding of the non-ideal 
MRI is incomplete, and absent 
higher resolution calculations the quoted thresholds for MRI activity in 
disks subject to Ohmic / Hall or ambipolar effects should be regarded 
as preliminary. However, even a definitive local determination of the conditions necessary 
for the non-ideal MRI to operate would address only part of the broader 
physical problem. The sustenance of MHD turbulence in a 
local simulation depends upon the strength and geometry of the field 
threading the volume, which in more complete disk models would be 
specified by external factors: either the initial conditions of disk formation 
or global magnetic linkage between different regions of the disk 
\citep{tout96,uzdensky08,sorathia10}. Global numerical simulations 
\citep{fromang06} that include non-ideal processes \citep{dzyurkevich10} 
are currently at an early stage of development, but will be important 
for studying the evolution of magnetic field both within and above the 
disk plane.

\subsubsection{Ionization and recombination processes}
\label{sec:3_ionize}

The absolute importance of the non-ideal MHD terms depends upon the 
conductivity of the gas, and hence on the ionization fraction, $x = n_e / n$ 
(e.g. Equation~\ref{eq:3_eta} for $\eta$). Collisional ionization of trace 
species, primarily potassium, suffices to yield $x \sim 10^{-12}$ at 
$T \simeq 10^3 \ {\rm K}$, so gas temperatures above this limit are 
likely enough to assure MRI activity. In $\alpha$ models of protoplanetary 
disks \citep{bell97}, a mid-plane temperature of $10^3 \ {\rm K}$ corresponds to a 
radius between $\approx 0.1 \ {\rm AU}$ (for a disk accreting at $10^{-9} \ 
M_\odot \ {\rm yr}^{-1}$, with $\alpha = 10^{-2}$), and $\approx 5 \ {\rm AU}$ 
(at $10^{-5} \ M_\odot \ {\rm yr}^{-1}$). Gas in the innermost disk, including  
that interacting directly with the stellar magnetosphere \citep{konigl91}, can thus be 
assumed to be MRI-active without recourse to further calculations.

At larger radii in the disk, the collisional ionization fraction is for all 
practical purposes zero, and $x$ must instead be computed by balancing 
non-thermal sources of ionization with recombinations. Although the 
non-thermal power required to maintain $x$ above the threshold for MRI activity is 
modest \citep{inutsuka05}, the only 
indubitable source of significant  ionization is disk 
irradiation by stellar X-rays, whose strength is well-determined observationally. 
In the Taurus star-forming region, for example, \citet{gudel07} find 
that the X-ray luminosity of detected sources scales with the stellar 
mass as,
\begin{equation}
 \log (L_X / {\rm erg \ s}^{-1}) = 1.54 \log (M_* / M_\odot) + 30.31,
\label{eq:3_Lx} 
\end{equation}
for masses in the range $0.1 \ M_\odot \lesssim M_* < 2 \ M_\odot$. 
The spectral properties are complex, but a simple representation, 
used in an analysis of Orion sources by  \citet{preibisch05}, combines a 
``cool" thermal plasma component at $T \approx 10^7 \ {\rm K}$, 
with a hotter component at $T \approx 3 \times 10^7 \ {\rm K}$. X-ray 
photoionization calculations \citep{glassgold97}, and Monte Carlo 
simulations \citep{igea99,ercolano08}, show that these 
spectra and luminosities imply enough hard X-ray photons to ionize 
significant disk columns to interesting levels. An analytic 
form for the ionization rate $\zeta$ has been given by 
\citet{turner08}, who find that the results of \citet{igea99} 
for a star emitting 5~keV thermal X-rays at $L_X = 2 \times 10^{30} \ {\rm erg \ s}^{-1}$ 
can be fit as,
\begin{equation}
 \zeta (\Delta \Sigma) = 2.6 \times 10^{-15} 
 \left( \frac{r}{1 \ {\rm AU}} \right)^{-2} \exp 
 \left[ -\Delta \Sigma / 8.0 \ {\rm g \ cm}^{-2} \right] \ {\rm s}^{-1},
\label{eq:3_turnerfit} 
\end{equation}
where $\Delta \Sigma$ is the column measured downward from the 
disk surface. This simple expression underestimates the ionization rate 
near the disk surface; a more detailed fit valid for lower column densities 
is provided by \citet{bai09}.

\begin{figure}[t!]
\includegraphics[width=\textwidth]{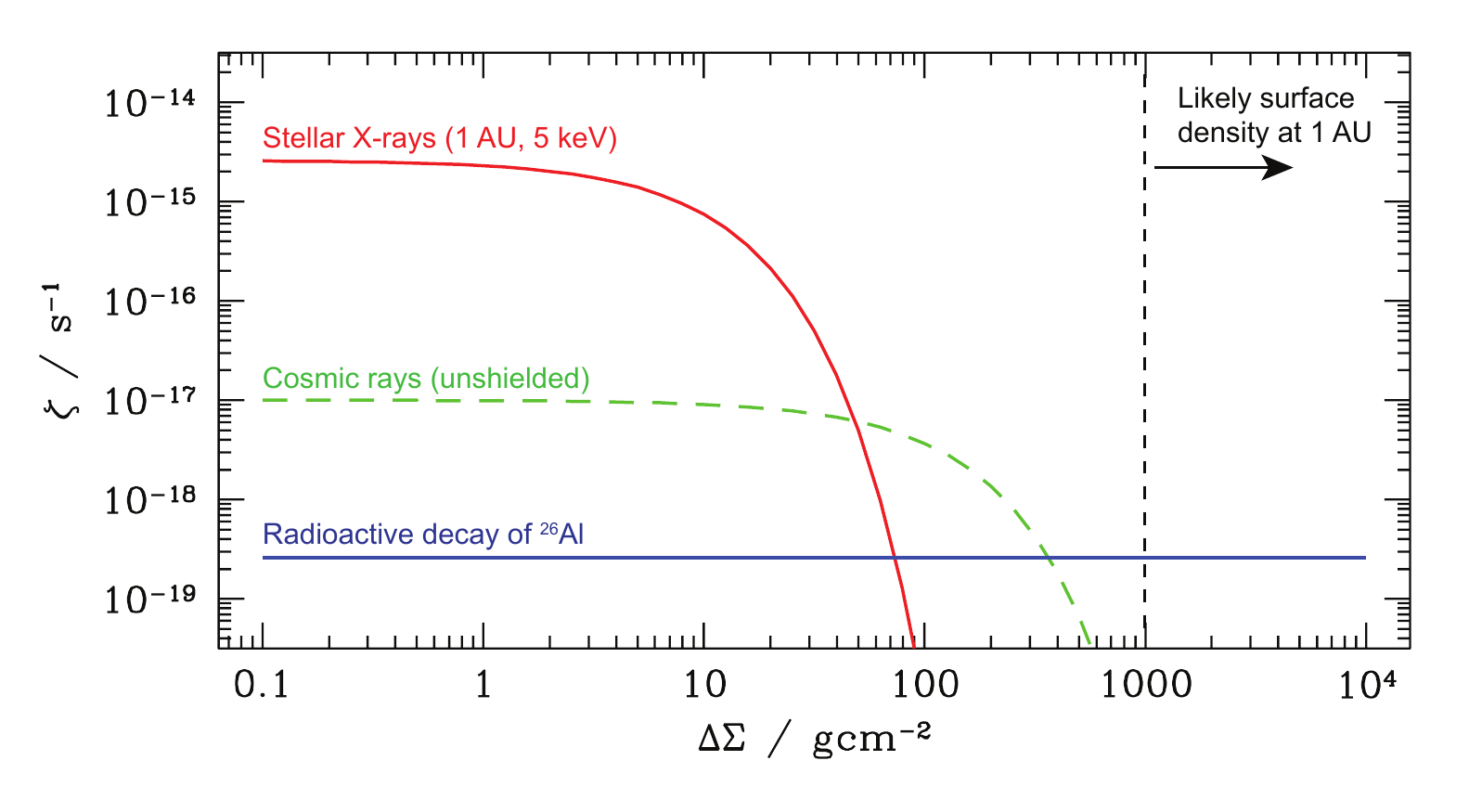}
\caption{The dependence of the ionization rate $\zeta$ on 
the column density $\Delta \Sigma$, measured from the disk surface. 
Contributions from the three 
three main sources of non-thermal ionization are shown: (a) stellar 
X-rays, evaluated at 1~AU (as a function of radius, $\zeta \propto r^{-2}$) 
using the \citet{turner08} fit to the 
calculations of \citet{igea99}, (b) cosmic rays, assuming that the 
flux is not attenuated by a stellar or disk wind, and (c) radioactive decay 
of the short-lived isotope $^{26}$Al \citep{stepinski92}. Although all three 
curves are approximate, they illustrate two key points. First, irradiation by 
stellar X-rays almost certainly dominates the ionization near the disk 
surface. Second, the likely column densities in the inner disk ($\Sigma \gtrsim 
10^3 \ {\rm g} \ {\rm cm}^{-2}$) are sufficient to provide effective shielding 
of the mid-plane from incident X-rays or cosmic rays.}
\label{fig:3_ionization}
\end{figure}

Figure~\ref{fig:3_ionization} shows the vertical dependence of $\zeta$ 
implied by Equation~(\ref{eq:3_turnerfit}), together with similarly rough 
estimates of the ionization rate that cosmic rays \citep{umebayashi81,gammie96} 
and radioactive decay of short-lived isotopes 
\citep[long-lived radionuclides can be neglected,][]{umebayashi09} might yield \citep{stepinski92}.  
\citep[][have proposed that energetic stellar protons may also 
contribute a significant steady flux]{turner09}. 
The level of these other non-thermal sources of ionization is more 
uncertain than for stellar X-rays. In the Solar System, the magnetic field in the 
Solar wind serves to partially exclude cosmic rays \citep{spitzer68}, and the 
presumably more powerful stellar and disk winds of T~Tauri stars may 
likewise shield the disk from the unattenuated cosmic ray flux. Similarly, isotopes 
such as $^{26}$Al may largely be locked up into grains. Fortunately, 
the qualitative features of the ionization rate profile are largely independent 
of these unknowns. Generically, stellar X-rays will be the dominant 
ionization source near the disk surface, while the 
columns expected in disks ar $r \sim 1 \ {\rm AU}$ are more than large 
enough to shield the mid-plane from any reasonable ionizing flux of 
photons or particles. Even the maximum plausible ionization rate due 
to radioactive decay is small.

These uncertainties in the ionization rate pale in comparison to 
those afflicting estimates of the recombination rate. The obstacle  
here stems not from the physics of recombination, which 
is reasonably well-understood \citep[for a discussion of the uncertainties 
due to disk chemistry, see e.g.][]{vasyunin08}, but rather from the fact that the 
answer depends sensitively on the unknown abundance of small dust grains. 
Before turning to this aspect of the problem, we first address the 
rate of recombination in dust-free gas. In this limit, the important 
reactions are dissociative recombination reaction with molecular 
ions, such as,
\begin{equation}
 {\rm HCO}^ + + e^{-} \rightarrow {\rm CO} + {\rm H},
\end{equation}
radiative recombinations with metal ions,
\begin{equation}
 {\rm Mg}^+ + e^{-} \rightarrow {\rm Mg} + \gamma,
\end{equation}
and reactions which transfer charge from molecular to metal 
ions,
\begin{equation}
  {\rm HCO}^ + + {\rm Mg} \rightarrow  {\rm HCO}  + {\rm Mg}^+.
\end{equation}
The simplest calculation of the recombination rate, due to 
\citet{oppenheimer74}, is based upon the equilibrium solution 
to a chemical network comprising ionization together with 
generic versions of the three reactions listed above. This 
network is simple enough that it can be solved analytically 
in some limits (e.g. Equation~\ref{eq:3_xactual}, if metal 
ions can be neglected), but it is already instructive. In 
particular, it reveals that the ionization fraction is sensitive 
to trace amounts of metal atoms. Their abundance is 
important because molecular ions (which would otherwise 
recombine efficiently with electrons), tend to transfer charge 
to the metals which then recombine relatively slowly 
\citep{fromang02}. This physical effect is also present in 
more refined calculations, which model reactions between 
large numbers of chemical species \citep{sano00,semenov04,illgner06,bai09}, 
though at a weaker level, because the more complex models 
include additional avenues for recombination of molecular ions. 
The issue is discussed in detail by 
\citet{illgner06}, who present a 
detailed comparison of the ionization fraction predicted by 
different reaction networks. 

Small dust grains, if they are present in any significant abundance, 
generally result in faster recombination. The collision rates between 
grains and charged particles have been computed by \citet{draine87}, 
who give expressions for recombination rates that include the effects of 
charged grains, and the induced polarization of neutral grains by electrons 
or ions \citep[a clear discussion in the disk context is given by][]{bai09}. 
These subtleties are important, but even a crude estimate suffices 
to demonstrate that small grains can radically alter the gas-phase recombination 
rate. Let us consider a disk with gas density $\rho$ and dust-to-gas ratio 
$f_{\rm d}$, in which the dust particles all have radius $a$ and material 
density $\rho_{\rm d}$. The surface area per unit volume presented by 
the dust particles is $S = 3 f_{\rm d} (\rho / \rho_{\rm d}) a^{-1}$, and the 
geometric estimate of the recombination rate due to collisions with molecular 
ions of number density 
$n_I$ and velocity $v_I \sim (kT / m_I)^{1/2}$ is $\sim S n_I v_I$. Comparing 
this to the rate of dissociative recombination, quoted in \S\ref{sec:3_intro}, 
we find that the ratio of the rate of recombination on dust compared to 
its gas-phase value is,
\begin{equation}
 \frac{\dot{n}_{I,dust}}{\dot{n}_{I,gas}} \sim 
 20 \left( \frac{f_{\rm d}}{10^{-2}} \right) 
 \left( \frac{x}{10^{-12}} \right)^{-1} 
 \left( \frac{T}{100 \ {\rm K}} \right)
 \left( \frac{a}{1 \ \mu {\rm m}} \right)^{-1}.
\end{equation} 
The conclusion, which is borne out by all detailed calculations, is that 
the ionization fraction in the non-thermal zone will be set largely by 
the abundance of grains \citep{sano00,illgner06,bai09}. What matters 
most is the size of the smallest grains, since for a standard grain 
size distribution \citep[$N(a) \propto a^{-3.5}$,][]{mathis77} the 
largest contribution to the surface area comes from the smallest 
particles. The limiting case is a size distribution that contains a 
significant abundance of polycyclic aromatic hydrocarbons (PAHs). 
These grains have sizes of just $a \sim 0.01 \ \mu {\rm m}$ 
and hence are especially efficient at stimulating recombination 
\citep{sano00,bai09,perezbecker10}.

The upshot of this discussion is that the ionization fraction within 
disks -- outside of the thermally ionized inner region -- is closely 
coupled to grain growth and grain settling. There is strong theoretical 
evidence that micron and sub-micron sized grains can coagulate 
rapidly, but this must be weighed against equally compelling 
observational evidence that at least some small particles survive 
throughout the disk lifetime \citep{dullemond05}. Settling can also 
be rapid, but will be inhibited for small particles if there is any 
significant level of turbulence within the gas \citep{dubrulle95,fromang06}. 
Since the level of turbulence is likely to depend, in turn, on the 
ionization fraction, it is easy to speculate that the chain of dependencies 
might give rise to either self-limiting or cyclic behavior, in which 
the dust settles until the ionization fraction rises enough to ignite 
the MRI.

\subsubsection{Layered accretion and dead zones}

\begin{figure}[t!]
\includegraphics[width=\textwidth]{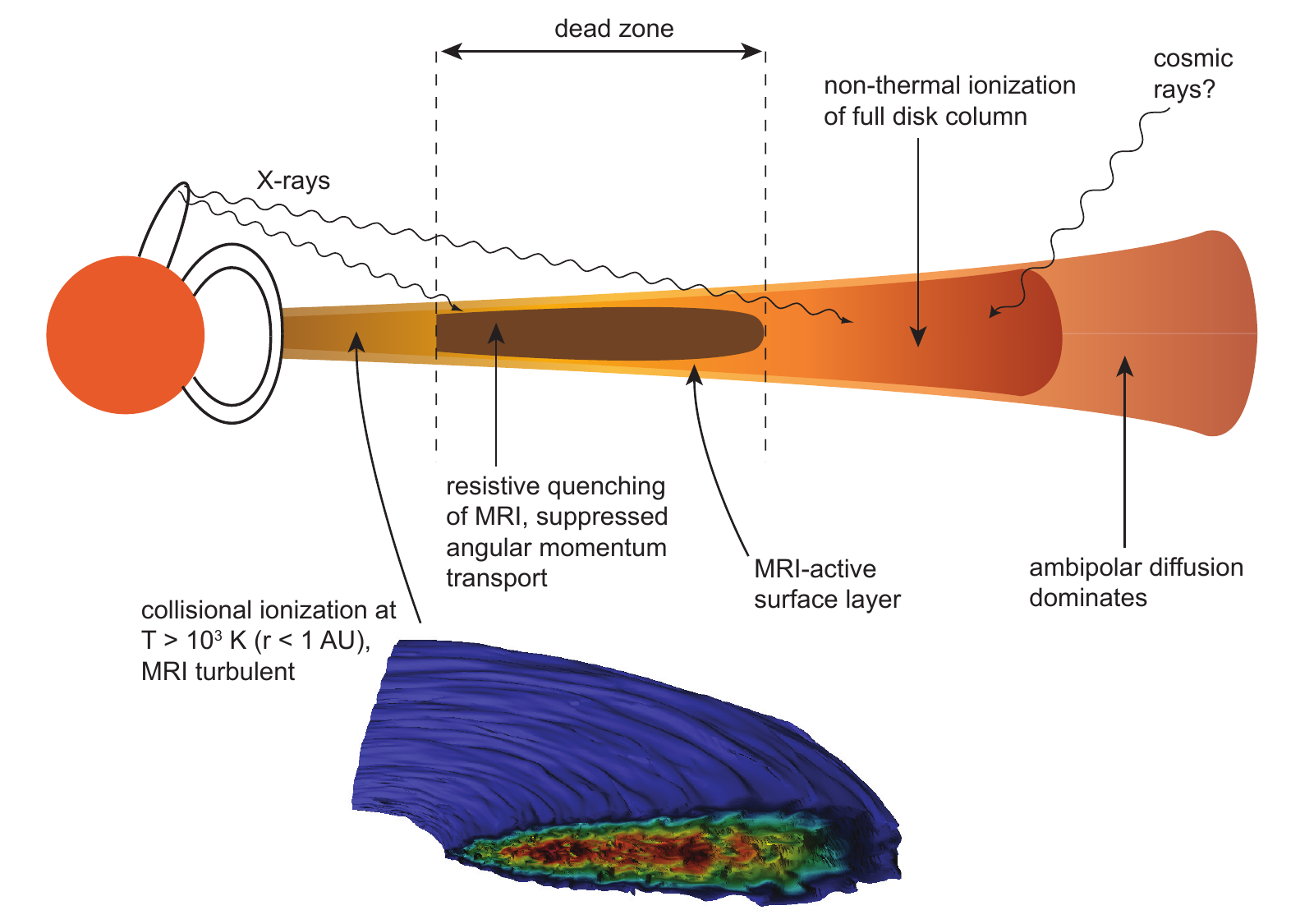}
\caption{Schematic structure of the protoplanetary disk if the low ionization 
fraction at radii $r \sim 1 \ {\rm AU}$ quenches angular momentum transport 
due to the MRI, forming a ``dead zone" \citep{gammie96}. X-rays, produced 
from the cooling of plasma confined within magnetic field loops in the 
stellar corona, ionize the disk surface, but fail to penetrate to the mid-plane. 
The inset figure shows density isosurfaces computed from a simulation of 
a fully turbulent disk \citep{beckwith11}.}
\label{fig:3_deadzone}
\end{figure}

An audacious synthesis of many of the ideas outlined above was 
attempted by \citet{gammie96}, who proposed that disks in which 
angular momentum transport occurred via the MRI ought to develop 
a ``layered" structure. The idea, illustrated in a somewhat modified form in 
Figure~\ref{fig:3_deadzone} \citep[with X-rays replacing cosmic 
rays as the main non-thermal ionization source, following][]{glassgold97}, 
is that the criteria for MRI activity define three radial zones,
\begin{enumerate}
\item
An inner zone, where the disk is MRI-active and turbulent because 
the mid-plane temperature is high enough ($T \gtrsim 10^3 \ {\rm K}$) to 
ionize the alkali metals. The outer boundary of this region, which is 
at about 1~AU for $\dot{M} = 10^{-7} \ M_\odot \ {\rm yr}^{-1}$ 
(Figure~\ref{fig:2_conditions}) will move 
inward as the disk accretion rate declines.
\item
An outer zone, where non-thermal sources of ionization suffice to 
raise the ionization fraction at the mid-plane above the threshold 
for MRI activity. The inner boundary of this region will also move 
inward as the surface density drops, and the shielding the disk 
provides to ionizing radiation and particles becomes less effective.
\item
An intermediate region, where the mid-plane is cool enough, and 
well-enough shielded from ionizing radiation, to fail to satisfy 
the conditions for the MRI to operate. \citet{gammie96} suggested that 
the disk at these radii would develop a layered structure, with a 
``dead zone" near the mid-plane in which turbulence was absent 
or strongly suppressed. Accretion 
would then occur entirely (or primarily) through an active surface 
layer, whose thickness is defined by the flux and penetration 
strength of cosmic rays (in the original version) or stellar X-rays.
\end{enumerate}
No observation provides direct evidence either for or against the 
existence of dead zones. Theoretical 
calculations, however, continue to suggest that it is more likely than not 
that protoplanetary disks develop a dead zone at radii 
$r \sim 1 \ {\rm AU}$ \citep{bai09,salmeron08,terquem08,turner09,turner10}. 
The sole situation in which a region of suppressed MRI transport is not 
predicted to exist is the case where there is a strong depletion of small 
grains. It is worth noting that, although all of these authors hew to the 
same basic considerations of non-ideal MHD physics, there are many 
significant differences between the technical approaches that they 
adopt. \citet{salmeron08}, for example, assume a minimum-mass 
model for the disk surface density profile \citep{hayashi81}, within which they compute 
the detailed vertical structure of  linear MRI modes \citep[see also][]{salmeron05}. 
\citet{terquem08}, on the other hand, adopts a simpler model for the 
MRI physics but solves self-consistently (in the context of a one-dimensional 
vertically averaged disk model) for the steady-state surface 
density distribution of the disk. Although the problem is undoubtedly 
messy, the general agreement between these calculations lends 
weight to the contention that some sort of dead zone does exist.

The existence of dead zones, falling as they do squarely in the midst 
of the radii of greatest interest for terrestrial planet formation, has 
multiple broader ramifications. The most important is probably the 
effect that a dead zone would have on planetesimal formation, since 
many candidate mechanisms for building planetesimals are sensitive 
to the level of intrinsic disk turbulence \citep{chiang10}. Generically, 
a dead zone also results in an enhanced surface density across the 
radii where the MRI is suppressed, and can -- if the suppression 
is complete enough -- admit new mechanisms for time-dependent 
disk outbursts \citep{gammie99,armitage01,zhu10b}. 

Going beyond the basic question of whether dead zones exist, 
it is important to know both the strength and the 
nature of any residual turbulence and angular momentum transport 
that persists in the MRI-inactive layer. Although the original models 
for layered accretion typically assumed entirely laminar, torque-free 
dead zones, subsequent work has identified three processes that 
likely animate the mid-plane to some degree. First, hydrodynamic 
waves, excited by the turbulence in the active surface layers, can 
penetrate to the mid-plane and exert a non-zero Reynolds stress 
on the magnetically inert gas there \citep{fleming03,oishi09}. In the 
vertically stratified local simulations of \citet{fleming03}, the 
mid-plane Reynolds stress in the dead zone was found to be 
surprisingly large -- less than an order of magnitude below the 
Maxwell stress in an MRI-active disk. Those simulations used 
an isothermal equation of state, and the true strength of Reynolds 
stresses communicated to the mid-plane may well be sensitive 
both to that assumption \citep{bai09} and to the thickness of 
the active surface layer. Nonetheless, a non-zero dead zone 
stress due to this effect seems very probable. \citep[Direct 
activation of the dead zone, due to turbulent mixing of ionized 
gas downward before it has time to recombine, has also been 
suggested, and can be effective in the absence of dust;][]{turner07,illgner08}. 
Second, radial magnetic fields can diffuse to the mid-plane, 
where shear suffices to generate an azimuthal component 
\citep{turner08}.  The combination of radial and azimuthal fields 
then yields a laminar Maxwell stress $\langle B_r B_\phi \rangle$, 
which can transport angular momentum (Equation~\ref{eq:2_stresses}) 
without, in principle, being accompanied by turbulent motions. 
This physics is related to that driving the time-dependent behavior observed by 
\citet{simon10} in moderately resistive disks (where the azimuthal 
field eventually grows strong enough to reignite the MRI even in the 
mid-plane), and leads one to suspect that the radial edges of dead 
zones may be significant wellsprings for accretion variability.
Finally, \citet{latter10} have identified a linear buoyancy-type instability 
that may operate in dead zones where resistive diffusion is much 
faster than thermal diffusion. This instability illustrates the general 
point that turbulence is not synonymous with angular momentum transport, 
since it would likely result in turbulence and mixing 
without significant attendant transport.

Thus far, work on MHD processes within protoplanetary disks has 
focused on dynamics. Crudely put, what is the value of $\alpha$, 
and how can that be related to observations of disk evolution? 
This emphasis should not blind us to the fact that 
other connections between observations and theory may ultimately be 
more powerful. \citet{king10}, for example, have emphasized that 
the energetic requirements for forming chondrules \citep[for a review, 
see][]{scott07} are substantial, 
and that meteoritic constraints from the Solar System can  
therefore inform models of accretion (whether by the MRI, gravitational 
instability, or some other process). A limited amount of work has 
already been done in this direction. \citet{joung04} suggested that 
the temperatures attained in current sheets could reach values high enough 
to form chondrules late in the disk evolution, when dust had 
settled and ambipolar diffusion was important \citep[the role of 
ambipolar diffusion in modifying the small-scale structure of MHD 
turbulence is discussed by][]{brandenburg94}. Additional work along 
these lines, including simulations in the Hall / Ohmic regime that is expected 
to prevail on AU scales at earlier times during disk evolution, would be 
valuable.

\subsection{Baroclinic instability and vortices}
The existence of dead zones in protoplanetary disks means that we 
have to worry about the effects of possible hydrodynamic instabilities, 
even if their angular momentum transport efficiency is low. In this 
sense protoplanetary disks are quite distinct from other astrophysical 
disk systems, in which purely hydrodynamic instabilities, even if they 
theoretically exist, would likely be swamped by the MRI. At the outset, 
it is useful to distinguish between instabilities that owe their origin 
purely to the shear (in the MHD case, the MRI is an example of such 
an instability), and those which instead feed off an unstable entropy 
gradient. For shear-driven instabilities, theoretical arguments \citep[e.g][]{lesur05} and, to a limited 
extent, laboratory experiments \citep{ji06}, have largely excluded 
the existence of instabilities that would yield efficient angular 
momentum transport. Purely hydrodynamic disks are linearly 
stable, and, although perturbations within such disks can 
exhibit transient growth \citep{ioannou01,afshordi05,chagelishvili03,rebusco09}, 
no avenue that leads to fully-fledged turbulence has been demonstrated. 

Entropy-driven instabilities are more promising. \citet{klahr03} observed 
an instability, which led to vortex formation and angular momentum transport, 
in global numerical simulations of disks with an outwardly declining entropy 
profile. These early simulations, however, were not able to clearly 
characterize the nature of the instability (for example, whether it was local or 
global), or the physical prerequisites for its existence. Subsequently, 
\citet{petersen07a,petersen07b}, using two-dimensional global simulations of 
unmagnetized disks, and \citet{lesur10}, using local simulations, confirmed 
that a disk with a linearly stable radial entropy gradient could nevertheless be 
unstable to the baroclinic generation of vorticity. The instability is known 
as the  ``subcritical baroclinic instability" (SBI).

Compared to the linear instabilities that anchor  
discussions of the MRI and disk self-gravity, the SBI is altogether 
more subtle. As currently understood, it is a non-linear 
(or ``subcritical") instability 
that relies for its existence on the interplay of two distinct thermodynamic 
properties of the disk, (1) a radially unstable entropy gradient and, (2)  
radiative cooling (or thermal diffusion) at a rate that promotes the baroclinic 
generation of vorticity \citep{petersen07a,lesur10}. The first of these properties is related to the 
stability of a rotating flow to convection, and can readily be quantified. 
The importance of buoyancy (in $r$) is expressed via the 
Brunt-V\"ais\"al\"a frequency, 
\begin{equation}
 N_r^2 = - \frac{1}{\gamma \rho} \frac{{\rm d}P}{{\rm d}r} 
 \frac{\rm d}{{\rm d}r} \ln \left( \frac{P}{\rho^\gamma} \right), 
\end{equation}
where $P = P(r)$ and $\rho = \rho(r)$ are the pressure and 
density profiles of the disk and $\gamma$ is the adiabatic index. 
In the absence of shear, the flow is unstable to the linear growth 
of perturbations if,
\begin{equation}
 N_r^2 < 0,
\end{equation}
which is the Schwarzschild criterion for the onset of convection. 
Including shear, the condition for 
linear instability to axisymmetric perturbations becomes the 
more stringent Solberg-Ho\"iland criterion, which can be 
written for a Keplerian disk as,
\begin{equation}
 N_r^2 + \Omega_{\rm K}^2 < 0.
\end{equation} 
For the SBI to operate; the background structure of the 
disk must be Schwarzschild unstable (but Solberg-Ho\"iland stable, 
thereby avoiding linear instability) in the radial direction. 

\begin{figure}[t!]
\center
\includegraphics[width=4.0in]{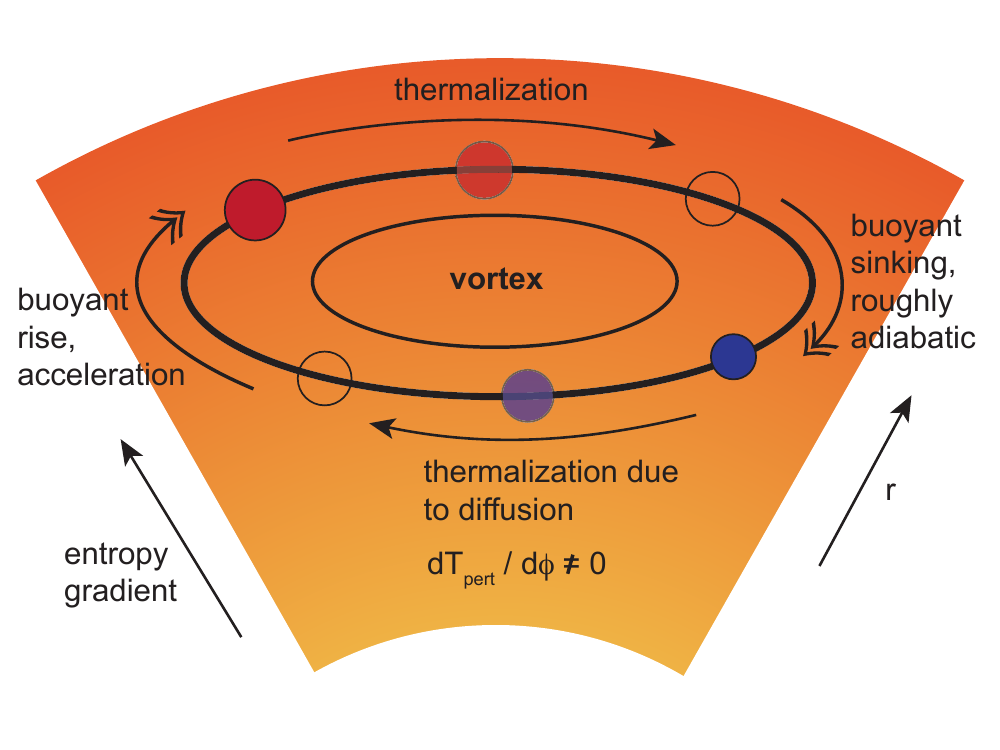}
\caption{Schematic illustration \citep[after][]{petersen07a,lesur10} showing 
how the baroclinic instability sustains non-linear vortices in a disk with 
a radial entropy profile. Fluid parcels moving either inward or outward 
are subject to buoyancy forces, and arrive at their new radial location 
either hotter or cooler than the background disk. As they drift azimuthally, 
diffusive processes equilibrate their temperature until it matches the 
background. As a consequence, the vortex sets up a temperature 
perturbation ${\rm d}T_{\rm pert} / {\rm d} \phi \neq 0$, maintaining the 
baroclinic term in the vorticity equation that drives the vortex.}
\label{fig:3_baroclinic}
\end{figure}

\citet{petersen07a} and \citet{lesur10} have studied the evolution 
of a Schwarzschild unstable disk that is initially seeded with finite 
amplitude temperature or vorticity perturbations. Vorticity is not 
a conserved quantity in such a setup, because there is a non-zero 
baroclinic term $\nabla P \times \nabla \rho$ that acts a source 
or sink of vorticity. Irrespective of the thermal physics of the disk, 
the flow will initially develop an organized pattern of vortices. In the 
absence of radiative cooling and heating, or of thermal diffusion, 
however, numerical simulations show that the vortices eventually 
decay. A self-sustaining, non-linear instability -- the SBI -- is only 
present when thermal or radiative diffusion occurs on an adroitly 
chosen time scale, neither too fast, which would reduce the 
effects of buoyancy, nor too slow, which would make the flow 
around the vortex almost adiabatic. Under the appropriate 
conditions, it is possible to set up a cycle, illustrated in 
Figure~\ref{fig:3_baroclinic}, in which dissipative processes are balanced by 
work done by buoyancy as fluid circulates around a self-sustaining 
vortex. 

Although the existence of the SBI in model systems is now 
securely established, we have only paltry knowledge the 
outcome of the instability, especially in three dimensions, 
where vertical stratification is sure to introduce new effects \citep{barranco05}.
The simulations of  \citet{lesur10} find that the instability 
is strongest for a diffusion time scale that is of the order of 
$10 \ \Omega_{\rm K}^{-1}$, and suggest that density waves 
generated by the vortices (observed in compressible simulations) 
may generate a weak outward flux of angular momentum 
\citep{johnson05}. It is also unclear under what physical conditions the 
SBI could play an important role in protoplanetary 
disks. For a disk to be Schwarzschild unstable, we require some 
combination of a relatively steep radial temperature profile, and a 
relatively shallow density profile. Specifically, if the mid-plane 
density and temperature scale as $\rho \propto r^{-\beta_\rho}$ 
and $T_c \propto r^{-\beta_T}$, respectively, then entropy 
declines outward if \citep{klahr03},
\begin{equation}
 \beta_T - (\gamma - 1) \beta_\rho > 0.
\end{equation} 
The temperature gradient at the mid-plane is likely to be steeper 
in disks whose thermal structure is dominated by viscous dissipation, 
rather than by external irradiation, and taken in isolation this 
suggests that the SBI might be most important in dense disks 
at early times, when accretional heating can in principle overwhelm 
irradiation. The requirement for a relatively short radiative cooling 
time scale, however, works in the opposite direction, since short 
cooling times occur for low mass disks of modest optical depth 
(e.g. Figure~\ref{fig:2_conditions}).
 
As with the MRI and self-gravity, interest in the SBI and disk vortices 
is not solely -- or in this case even primarily -- 
motivated by dynamical considerations. Vortices  
concentrate solid particles toward their cores \citep{barge95,tanga96,chavanis00}, 
which as a result become sites of locally enhanced solid-to-gas ratio. The 
existence of vortices could therefore have profound implications for 
planetesimal formation, particularly if planetesimals form via a mechanism 
that requires a threshold solid-to-gas ratio to be exceeded \citep[an example being 
the streaming instability;][]{johansen09}.

\subsection{Convection}
Convection has probably generated more vigorous instability among 
theorists than it does in typical protoplanetary disks. Early disk models, 
notably that of \citet{lin80}, were based upon the assumption that convective cells, with   
characteristic velocity $v_c$ and scale $l \sim v_c / \Omega_{\rm K}$, 
would transport angular momentum with an effective kinematic 
viscosity $\nu = v_c^2 / \Omega_{\rm K}$. Standard mixing length 
theory was used to calculate $v_c$ as a function of the vertical 
disk structure. These models were later abandoned in the face of 
persuasive numerical and analytic evidence suggesting that 
vertical disk convection was not merely an inefficient source of 
viscosity, but actually worked to transport angular momentum 
in the wrong direction, inward! Very recently, however, sentiment 
has reversed sign again. \citet{lesur10b} have completed high 
resolution incompressible simulations of disk convection 
that show outward transport of angular momentum, with 
efficiencies that under reasonable disk conditions might 
yield $\alpha \sim 10^{-4}$ or even larger. The contrary 
results of previous investigators can be traced to numerical 
shortcomings (early simulations were not able to access the  
strongly turbulent regime of convection that ought to prevail in 
real disks), and to the use of analytic 
approximations that prove to be invalid. 

The new results are important as a demonstration of how a 
fundamental physical process operates in a disk geometry, 
but it is as yet unclear whether they presage a 
revival of disk models in which convection 
plays a leading role. Convection cannot be important in an 
externally irradiated disk, and the 
current idealized simulations do not model the viscous dissipation that, in  
a self-consistent theoretical description, would have to sustain the unstable 
stratification. Moreover, 
although \citet{lesur10b} obtain outward 
transport, they also confirm the intuitive expectation that convective 
turbulence is more adept at moving heat (vertically) than 
it is at moving angular momentum (radially). This observation 
certainly limits the efficiency of a convectively-driven disk, but, 
until more work has been done to confirm the new results and 
establish the fundamental 
properties of disk convection, more definitive statements are 
speculative.

\subsection{Planet-driven disk evolution}
Theoretical estimates of the time needed to assemble the core of a 
gas giant planet \citep{movshovitz10} are, to within uncertain factors 
of a few, equal to the observed lifetime of protoplanetary gas disks. 
Several authors have speculated that this apparent coincidence might, 
to the contrary, signal a causal relationship; perhaps the formation 
of planets itself catalyzes angular momentum transport and results 
in the viscous dispersal of the gas disk \citep{goodman01,sari04}. 
The physical process that can mediate such an unexpected coupling 
is well known. Gravitational torques between a planet and a gas 
disk, exerted at radial locations within the disk that correspond to 
Lindblad resonances, transfer angular momentum outward in a 
manner qualitatively resembling a viscous process \citep{goldreich80}.

In its simplest version, the idea that planets generate the bulk of 
angular momentum transport within disks almost certainly fails. 
Although, in principle, a low-mass planet embedded within the  
gas disk could passively shuttle angular momentum between the 
disk at its inner and outer Lindblad resonances, while remaining 
on a fixed orbit, no calculation suggests that gravitational 
torques operate in this fashion. Rather, calculations show that 
the net torque on the planet (due to the sum of Lindblad and 
corotation torques) is a significant fraction of the sum of the 
absolute values of the torques \citep[this is the well known problem of 
Type~I migration, e.g.][]{lubow10}. As a result, a population of 
low mass planets would only be able to catalyze the accretion 
of a relatively small mass of gas -- of the order of the total mass 
in planets -- before either spiraling into the star or away to large 
disk radii.

Nonetheless planets -- especially massive gas giants -- may 
still play a significant if more limited role in disk evolution. 
The interpretation of transition disk sources, observational 
aspects of which are reviewed by \citet{williams11} in this 
volume, remains unclear, but it is undisputed that the 
existence of large inner holes in transition disks cannot be explained 
by simple viscous processes acting within the gas. A single 
massive planet, which opens a gap in the gas disk via the action 
of gravitational torques while allowing some gas to flow 
past the planet toward the star, could account for the observed 
properties of some transition disk sources \citep{lubow06,rice03b,calvet02}. 
In other instances, however, the overlapping gravitational torques 
from multiple gas giant planets \citep[][Z. Zhu, private communication]{morbidelli07}, 
acting in concert more like a 
viscous process, may be needed if one is to explain the data 
within a planetary scenario.

\subsection{Section summary}
Determining the strength and nature of angular momentum transport is 
particularly difficult for protoplanetary disks, since no single physical 
mechanism is clearly dominant.  A 
robust theoretical determination of $\alpha(r,z,t)$ is not possible, but 
enough is known about candidate transport 
mechanisms to make predictions that could be tested with future disk 
observations. To summarize this Section:
\begin{itemize}
\item
Disk self-gravity (at early times), and the magnetorotational instability, 
are likely the most important dynamical agents in viscous protoplanetary 
disks. Other processes can also transport angular momentum, but have 
less general applicability.
\item
The efficiency of the MRI is a partially-known function of the magnitude of 
non-ideal MHD terms. In a real disk, these will vary with the stellar 
properties (since X-ray ionization is an important process), and with the 
extent of dust coagulation (since small dust particles largely determine the recombination rate). 
Models that include the non-ideal physics are uncertain, but most of them predict 
the existence of a dead zone, with reduced levels of turbulence, at radii 
of the order of an AU.
\item
The possibility that magnetic braking dominates angular momentum loss from 
some disks must be kept in mind. Even if it does not, weak large-scale magnetic 
fields threading the disk affect the efficiency of MRI-driven transport.
\end{itemize}

\section{DIFFUSION AND MIXING}
Thus far, we have proceeded under the assumption that angular momentum 
transport within the disk derives from a turbulent process. Turbulence is 
also, generically, efficient at mixing different fluids together, so if our 
working assumption is valid there ought to be a close connection between 
angular momentum transport, disk evolution, and the diffusion of trace gas species 
and dust through the disk. To quantify that connection, we assume that the 
disk contains some dynamically unimportant trace species of gas, with 
density $\bar{\rho}$, that mixes vertically according to a standard diffusion 
equation,
\begin{equation}
 \frac{\partial {\bar{\rho}}}{\partial t} = 
 \frac{\partial}{\partial z} 
 \left[ D \rho \frac{\partial}{\partial z} 
 \left( \frac{\bar{\rho}}{\rho} \right) \right],
\end{equation}
where $D$, the vertical gas diffusion coefficient, has the same units as 
viscosity (cm$^2$~s$^{-1}$). The relative efficiency with which the 
turbulence transports angular momentum, and mixes gas vertically, can 
then be measured via the Schmidt number,
\begin{equation}
 {\rm Sc} = \frac{\nu}{D}.
\end{equation}
A Schmidt number ${\rm Sc} > 1$ thus means that the turbulence 
is relatively poor at mixing for a given angular momentum transport 
efficiency. Very similar considerations apply to radial diffusion, 
although for the radial case the concentration ${\bar{\rho}} / \rho$ 
must be obtained from the solution to an advection-diffusion equation  
that accounts for the radial flow of the background disk \citep{clarke88}.

The vertical Schmidt number has been measured for trace gas 
species (and for small dust particles, which when tightly coupled to the 
gas by aerodynamic forces behave similarly) using both local and global MRI 
simulations \citep{carballido05,illgner08,fromang09,turner06,johansen05}. 
The dimensional argument that $D \sim \nu$ turns out to be reasonable; 
typical values of the vertical Schmidt number extracted from simulations 
are ${\rm Sc} = 1 - 3$. Significant vertical variations occur -- unsurprisingly since the 
properties of fluid turbulence driven by the MRI vary markedly with 
height above the mid-plane -- and must be included in order to model 
the vertical distribution of diffusing species accurately \citep{fromang09}. 
Substantially higher Schmidt numbers result if the turbulence within 
the disk is stimulated by the presence of net vertical fields \citep{johansen06}.

Larger particles, that are imperfectly coupled aerodynamically to the 
turbulent motions within the gas, are expected to diffuse less readily 
than gaseous contaminants \citep{youdin07,cuzzi93}. The ratio of the 
particle diffusion coefficient, $D_p$, to that of the gas, $D$, can be 
estimated analytically using a model for the structure of the turbulence 
within the disk. It depends upon the strength of the coupling between the particles 
and the turbulent gas, measured via the dimensionless stopping time,
\begin{equation}
 \tau_s = \Omega_{\rm K} t_s,
\end{equation}
where $t_s$ is the exponential decay time for damping of relative motion 
between the particle and the gas. With this definition, \citet{youdin07} 
obtain,
\begin{equation}
 \frac{D_p}{D} \sim (1 + \tau_s^2)^{-1}.
\end{equation}
The diffusion of particles with dimensionless stopping times exceeding 
unity is thus predicted to be severely curtailed. Although variation of 
$\nu / D_p$ with stopping time (in the expected sense) has been seen 
in studies of particle diffusion within 
MRI turbulent disks \citep{fromang09}, an accurate test of this scaling 
(which differs from previously widely used results) is not yet available.

\section{PHOTOEVAPORATION}
\label{sec:5}
The observation of accretion signatures, generated when infalling gas strikes 
the stellar photosphere \citep{calvet98}, demonstrates that angular momentum 
transport or loss processes in the disk do occur on astronomically relevant 
time scales. This does not mean, however, that angular momentum transport processes 
are necessarily the primary agents for protoplanetary disk evolution. Mass loss from the disk 
is also likely to occur, perhaps especially at moderate to large distances from the 
star where the gravitational potential well is not too deep. If the mass loss rate 
is significant enough -- roughly speaking if it exceeds the accretion rate $\dot{M}$ -- 
then it will be the physics of mass loss rather than that of angular momentum 
transport that largely determines properties of the disk such as the 
surface density profile. This possibility needs to be kept in 
mind when interpreting observations of disks on large scales, 
where the local time scale for mass loss is plausibly shorter than that for 
angular momentum transport.

Here, we review work on disk mass loss driven by 
photoevaporation \citep{bally82}. The importance of photoevaporation 
depends upon the strength and spectrum of radiation incident on 
the disk -- either from the central star or from neighboring stars, if they are 
luminous enough as to dominate the flux of energetic photons -- 
and upon the composition of the disk's upper layers. Although estimates 
of the resultant mass loss rate are uncertain, photoevaporation is 
very likely an important process for many disks, and may dominate the 
evolution in the Classical T~Tauri phase for low mass stars in rich clusters, and 
for isolated stars that are particularly strong X-ray sources.

\subsubsection{Essential physics}

\begin{figure}[t!]
\includegraphics[width=\textwidth]{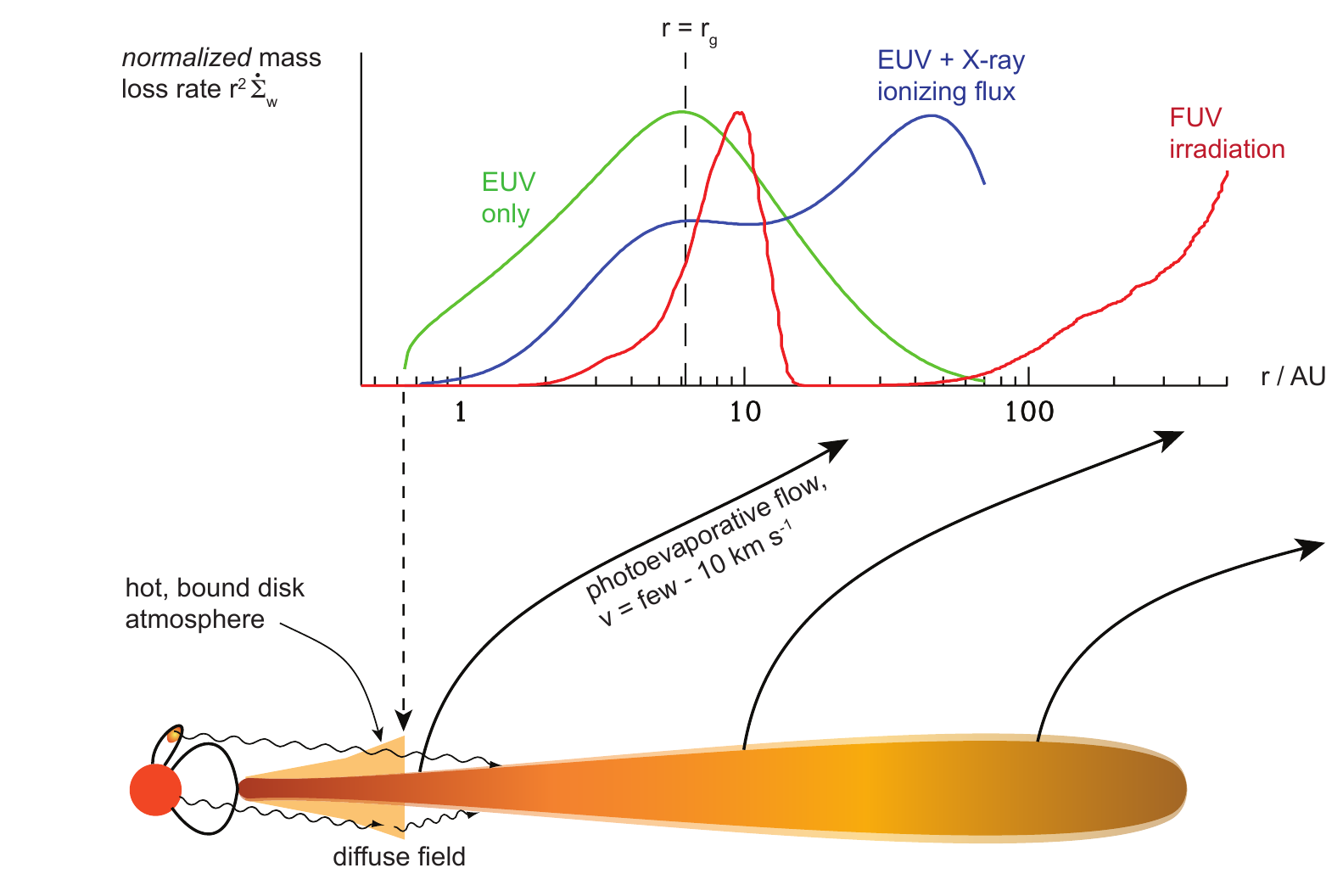}
\caption{Mass loss by photoevaporation occurs when FUV, EUV or X-ray radiation 
heats the disk surface to a temperature such that the gas is unbound. The normalized 
mass loss rate $r^2 \dot{\Sigma}_w$ is shown as a function of radius for three models: 
(1) EUV only, using an analytic fit to simulation results by \citet{font04} for $M_* = 0.7 \ M_\odot$, 
(2) an EUV plus X-ray model \citep{owen10}, also for $M_* = 0.7 \ M_\odot$, 
and (3) an FUV model for $M_*=M_\odot$ \citep[using data smoothed from 
Figure~5 of][]{gorti09b}. The EUV component 
drives mass loss primarily near a characteristic radius $r \sim r_g = GM_* / c_s^2$ 
(where $c_s$ is the sound speed in the ionized gas), while X-rays, produced in a 
magnetized stellar corona, or FUV 
irradiation produce a component of mass loss weighted to larger radii. 
The absolute mass loss rates depend upon the stellar spectrum. Note that 
different assumptions and computational methods were used to produce each 
of the three mass loss curves, which are therefore illustrative rather than strictly 
comparable.}
\label{fig:4_photoevaporation}
\end{figure}

The essential physics of photoevaporation is illustrated in Figure~\ref{fig:4_photoevaporation} 
\citep[after][]{shu93}. Ionizing or dissociating radiation from the central star, or from 
external stars, impinges on the upper and lower surfaces of the disk and heats a 
relatively thin skin of  gas there to a temperature $T_{\rm surf} > T_{\rm eff}$. 
At some radius $r_g$, the sound speed $c_s$ in the surface layer equals  
the local Keplerian velocity, where $r_g$ is given by,
\begin{equation}
 r_g = \frac{GM_*}{c_s^2}.
\label{eq:5_rg} 
\end{equation}
At radii $r \gtrsim r_g$ the gas in the surface layer 
is unbound (even neglecting the effects of rotation), and there is no impediment 
to it flowing freely away from the disk as a wind driven by thermal pressure gradients. 
For gas with mean molecular weight $\mu$, that has a number density at the base of 
the heated layer $n$,  dimensional arguments suggest that the local mass loss rate ought 
to be,
\begin{equation}
 {\dot{\Sigma}_{\rm wind}} (r) \sim \mu n(r) c_s(r).
\label{eq:5_sigwind} 
\end{equation}
Magnetic fields may be present, but need play no role at all in the driving of the 
wind. If they can be ignored, then the specific angular momentum of gas in 
the wind matches that at the surface of the disk where each outflowing 
streamline originates. 
In its simplest form, then, photoevaporation is a ``pure" mass loss process, that 
diminishes $\Sigma$ locally at each radius without causing additional evolution 
due to angular momentum loss.

The radial dependence of the disk mass loss rate is determined by the interplay of radiation physics 
(the spectrum of the irradiating sources, and radiative transfer effects 
that alter the distribution of the flux seen by the disk), thermal physics 
in the disk (which will determine $T_{\rm surf}$), and hydrodynamics 
(which finally determines $\dot{\Sigma}_{\rm wind}$). Photoevaporation 
models can be divided into two classes depending upon the 
origin of the disk-irradiating flux. ``External" photoevaporation models 
apply when the radiation that drives mass loss originates 
from other stars, usually massive stars in young clusters that have 
prodigious ultraviolet luminosities \citep{adams04,johnstone98}. On average, 
the importance of external 
photoevaporation increases with the mass and compactness of the stellar 
cluster within which a disk finds itself \citep{armitage00,fatuzzo08}. For 
disks around stars forming in groups or small clusters, such as Taurus, 
mass loss must instead be driven internally, by radiation from the 
central star. (Even in relatively rich clusters, internal photoevaporation can 
be the most important process at $r \sim \ {\rm AU}$, due to the 
$1/r^2$ dependence of the central stellar flux.) 

\subsubsection{X-ray and UV-driven photoevaporation}
At 100~AU, the simple argument given above (leading to Equation~\ref{eq:5_rg}) 
would imply that the surface layer of the disk must be 
heated to $T \gtrsim 10^3 \ {\rm K}$ in order to initiate a photoevaporative 
flow. This is in fact unduly pessimistic; more detailed hydrodynamic models show 
that gas can escape at a reduced rate from radii as 
small as $r \simeq 0.1 \ r_g$ \citep{begelman83,liffman03,font04}. 
Even with this boost, however, photoevaporation requires surface temperatures 
high enough that they can only be attained as a result of disk irradiation 
by energetic photons,
\begin{enumerate}
\item
Far-ultraviolet (FUV) radiation, with $6 \ {\rm eV} < h \nu < 13.6 \ {\rm eV}$, 
that is able to dissociate hydrogen molecules but not ionize hydrogen atoms.
\item
Extreme-ultraviolet (EUV) radiation, with $13.6 \ {\rm eV} < h \nu < 100 \ {\rm eV}$, 
capable of ionizing hydrogen.
\item
X-rays, defined by convention as photons with $h \nu > 0.1 \ {\rm keV}$.
\end{enumerate}
The rate of mass loss depends upon the 
original spectrum of the radiation, upon how much attenuation 
takes place along the line of sight to the disk surface, and upon 
how each type of radiation interacts and heats the disk surface.

The central region of the Orion nebula provides a good environment 
in which to test  external photoevaporation models, since both the 
relevant irradiating flux (FUV and EUV radiation from massive stars, 
especially the O6 star $\theta^{\rm 1}$~Ori~C), and the structure of the 
resulting photoevaporative flows, can be reliably inferred or directly 
observed \citep{odell93}. The Orion proplyd LV~2, for example, which is seen in 
projection to lie within the bright stars of the Trapezium, has a 
measured photoevaporative mass loss rate of  
$\dot{M}_{\rm wind} \simeq 6 \times 10^{-7} \ M_\odot \ {\rm yr}^{-1}$, 
with an estimated error of only $\sim$10\% \citep{henney02,vasco05}. 
The observations in Orion are generally found to be consistent with theoretical models 
for external photoevaporation \citep{adams04,johnstone98,clarke07,storzer99,richling00}, 
which show that such high rates of mass loss occur when large ($\sim 10^2$~AU)
and massive disks are placed in close proximity to massive stars. 
The dominant driver of photoevaporation under these conditions is 
FUV radiation, which heats the disk surface to a temperature that ranges 
between a few$\times 10^2 \ {\rm K}$ (for a modest flux of incident FUV, 
specified as $G_0 = 300$, where $G_0 = 1$ corresponds to a flux of $1.6 \times 10^{-3} \ {\rm erg} \ 
{\rm cm}^{-2} \ {\rm s}^{-1}$ between 91.2~nm and 200~nm), up to 
$10^3 \ {\rm K}$ for $G_0 = 3 \times 10^4$ \citep{adams04}. For disks 
that are large enough, $r \gtrsim 0.1-0.2 r_g$, the FUV-driven flow is 
dense enough as to completely absorb EUV radiation, whose strength 
and effects are therefore immaterial. Over time, the outer radius of the 
disk shrinks and the mass loss rate drops, precipitously once 
$r_{\rm out} \ll r_g$. The main effect of external photoevaporation 
is thus not to destroy disks entirely (though it does curtail their 
lifetime), but rather to truncate their sizes. For sufficiently 
strong FUV fluxes ($G_0 \gtrsim 10^4$) the truncation radius 
of 0.1-0.2~$r_g$ corresponds to radii of 10-20~AU, which is 
small enough to significantly limit the formation  
of gas giant planets in rich clusters.

Internal photoevaporation models are more complex, because a 
disk around a low mass star may be significantly influenced by 
radiation in all three bands: FUV, EUV, and X-ray. The flux 
seen by the disk in these bands is moderately well determined. 
In the FUV, there is likely to be both a persistent chromospheric 
contribution, estimated by \citet{gorti09} at 
$L_{\rm FUV} / L_* \approx 5 \times 10^{-4}$, and a larger 
component associated with accretion hot spots. An accretion 
rate of $\dot{M} = 10^{-7} \ M_\odot \ {\rm yr}^{-1}$, for example, 
could yield an FUV luminosity of the order of $10^{32} \ 
{\rm erg} \ {\rm s}^{-1}$. X-ray luminosities 
$L_X \approx 2 \times 10^{30} \ {\rm erg} \ {\rm s}^{-1}$ 
(for Solar mass stars, Equation~\ref{eq:3_Lx}) are comparable 
to the chromospheric FUV values, though it should be noted 
that the dispersion in X-ray luminosity is large. EUV 
luminosities are more problematic, because absorption 
in the interstellar medium frustrates direct observation 
of the EUV flux from accreting, Classical T Tauri stars (CTTs). 
Estimates in this band are generally derived using scaling 
arguments \citep{kamp04},  or indirect observational constraints 
\citep{alexander05}. For example, theoretical spectra, constructed assuming 
that the coronal properties of CTTs resemble 
those of RS CVn binaries (some of which lie near enough 
that their EUV is not completely absorbed), suggest that 
$L_{\rm EUV} \sim L_X$ \citep{ercolano09}. For reference, the kinetic 
luminosity of even a powerful wind -- one with a mass loss 
rate of $10^{-8} \ M_\odot \ {\rm yr}^{-1}$ and a terminal velocity 
of 5~km~s$^{-1}$ -- is only $L \approx 10^{29} \ {\rm erg} \ 
{\rm s}^{-1}$. Simple energetic arguments cannot therefore exclude 
any of the different radiation sources from playing a major role 
in disk mass loss. Observational 
evidence for photoevaporation driven by the central star is 
for now primarily indirect, being based, for example, on the 
comparison between observed [Ne II]~12.81${\mu \rm m}$ line 
profiles \citep{herczeg08,pascucci09} and 
theoretical models \citep{alexander08,ercolano10}.

Historically, photoevaporation driven by an EUV flux from the star 
received the first detailed study. The initial motivation was to 
explain the formation of ultracompact HII regions as a result of photoevaporation  
of the disk of a massive star \citep{hollenbach93,yorke93}, but it was 
shortly realized that similar processes could be significant for disks 
around low mass stars \citep{shu93}. The physics in this case is 
greatly simplified by the fact that the temperature of photoionized 
gas is nearly a constant, $T_{\rm surf} \approx 
10^4 \ {\rm K}$ ($c_s \approx 10 \ {\rm km} \ {\rm s}^{-1}$), independent of 
uncertain aspects of the disk's 
chemical properties or of dust physics. Provided that the EUV is 
not absorbed close to the star (as may happen early on), the 
ionizing radiation produces a modest but securely determined 
rate of mass loss. For a disk that extends to close to the star 
(i.e. before any inner hole develops), the mass loss rate can 
be evaluated by adopting the analytic scaling of  
\citet{hollenbach94}, scaled down by a factor of three as suggested 
by the numerical results of \citet{font04},
\begin{equation}
 \dot{M}_{\rm wind} \approx 1.4 \times 10^{-10} 
 \left( \frac{\Phi}{10^{41} \ {\rm s}^{-1}} \right)^{1/2} 
 \left( \frac{M_*}{M_\odot} \right)^{1/2} 
 \ M_\odot \ {\rm yr}^{-1}.
\end{equation}
Here, $\Phi$, the stellar output of ionizing EUV photons, 
affects the mass loss rate only as the square root. This  relatively 
weak scaling with $\Phi$, which is not qualitatively altered even if the wind 
contains dust \citep{richling97}, ameliorates  
somewhat the large uncertainties attending its correct value. The 
radial distribution of the mass loss is plotted in Figure~\ref{fig:4_photoevaporation}. 
EUV photoevaporation is most important across a relatively small range 
of radii, and peaks near $r_g$, at about 9~AU for a Solar mass star.

The relatively low rates of mass loss due to EUV photoevaporation can 
be attributed, in part, to the large cross section for absorption of ionizing 
photons by neutral gas. For the geometry shown in Figure~\ref{fig:4_photoevaporation} 
most EUV photons emitted by the star are absorbed by the tightly bound 
disk gas nearest the star, and the flux seen by the gas near $r_g$ is 
dominated by the diffuse field of photons produced as the inner gas 
recombines. Higher mass loss rates occur after the disk close to the star becomes optically thin to ionizing 
radiation, and the stellar flux can impinge directly upon the outer disk. 
In this ``direct" regime of EUV photoevaporation, 
the mass loss rate is predicted to increase with the radius of the 
inner hole in the disk as $r_{\rm hole}^{1/2}$ \citep{alexander06}.

Recent work shows that the addition of the X-ray and / or FUV spectral 
components leads to much higher predicted mass loss rates \citep{owen10,gorti09}. 
Both X-rays and FUV photons are more penetrating than EUV, and, although 
they do not heat the disk surface to as high a temperature as EUV, the 
boost in mass loss derived from launching the flow at a higher $n$ 
(Equation~\ref{eq:5_sigwind}) has the potential to outweigh the lower $c_s$. 
These flows, however, are less forgiving of analytic approximations 
than is the case with EUV irradiation, which generates a sharply defined 
ionization front that serves as the launch point for the wind. For X-ray 
or FUV irradiation, it is necessary to solve numerically for both the 
hydrodynamic and thermal structure of the disk surface as it transitions 
into a wind. The thermal structure  depends 
upon the irradiating spectrum, grain physics, chemical processes, 
and cooling by atomic and (where molecules are present) molecular lines 
\citep[e.g.][]{gorti08,woitke09}. Accounting accurately for these effects 
remains challenging, with independent codes returning somewhat different 
predictions for the gas temperature even when the same physical 
assumptions are adopted \citep{rollig07}. As was the case for the 
ionization fraction (\S\ref{sec:3_ionize})  the abundance of PAHs and 
small grains is critical. If PAHs are present, \citet{gorti08} find that their contribution to 
disk heating (via grain photoelectric 
emission) can be larger by a factor of $\sim$2 than X-ray heating at 
$r \sim 10$~AU. Observationally, PAH spectral features are 
detected in only a small fraction of T Tauri stars \citep{geers06,oliveira10}. 
The non-detections could, however, be explained by relatively modest 
depletion of the surface PAH abundance \citep[possibly as little as a 
factor of $\sim 10$;][]{geers06}, rather than by a complete absence of 
very small grains. This uncertainty in the PAH abundance (and in their properties) 
evidently propagates through to FUV photoevaporation models.

\citet{owen10} computed radiation hydrodynamic models of X-ray 
irradiated disks, assuming a luminosity 
$L_X = 2 \times 10^{30} \ {\rm erg} \ {\rm s}^{-1}$. The input spectrum 
included an EUV component ($L_{\rm EUV} = L_X$), while FUV 
irradiation was not considered. For a disk of mass $M_{\rm disk} = 0.026 
\ M_\odot$, surrounding a $0.7 \ M_\odot$ star, they obtained a mass 
loss rate of $\dot{M}_{\rm wind} = 1.4 \times 10^{-8} \ M_\odot \ {\rm yr}^{-1}$, 
some two orders of magnitude greater than the EUV-only expectation. 
Absolute mass loss rates were locally higher than the EUV models at 
essentially all radii, but, when normalized to the total mass loss rate, 
the X-ray wind was found to originate across a much broader range 
of disk radii, with the strongest contribution coming at large ($\approx 
50 \ {\rm AU}$, see Figure~\ref{fig:4_photoevaporation}) distances from the star.
Even larger mass loss rates from the outer disk are likely in the presence 
of strong FUV irradiation. \citet{gorti09} calculated the thermal and 
chemical structure of a disk exposed to the joint effects of FUV, EUV, and 
X-ray irradiation, from which they analytically estimated the likely 
mass loss rates. For an FUV luminosity of $L_{\rm FUV} = 4 \times 10^{31} 
\ {\rm erg} \ {\rm s}^{-1}$ incident on a disk around a Solar-mass star, 
they estimated mass loss rates of $\sim 3 \times 10^{-8} \ M_\odot \ 
{\rm yr}^{-1}$, with much of the wind originating beyond 100~AU.  

\subsubsection{Disk evolution}
Following the work of \citet{clarke01}, many authors have investigated 
how disks evolve under the combined influence of viscous angular 
momentum transport and photoevaporation 
\citep{alexander06b,adams04,matsuyama03,owen10,gorti09b}, 
and how such models may be constrained by observations of 
transition disk sources \citep{owen10b} and by the statistics 
of extrasolar planets \citep{alexander09}. The basic approach 
is to couple simple viscous disk models, that assume 
either a constant $\alpha$ or a fixed functional form for $\nu(r)$, 
with mass-loss profiles $\dot{\Sigma} (r)$ derived from analytic 
or hydrodynamic photoevaporation calculations. The wind is 
assumed to carry away the same specific angular momentum 
as the disk at the launch locations. Given the limited 
data against which these models can be tested, the simple-minded 
treatment of disk evolution is a virtue.

The inclusion of photoevaporation into disk models does not materially alter the 
derived observational constraints on the efficiency of disk angular momentum 
transport. Acceptable values of $\alpha$ derived by various 
authors range from $\alpha = 2.5 \times 10^{-3}$ \citep{owen10b} 
to $\alpha = 10^{-2}$ \citep{gorti09b}, no different from earlier studies 
that ignored mass loss \citep{hartmann98,hueso05}. Where photoevaporation 
matters -- and can potentially be tested -- is as an agent for disk dispersal. 
\citet{clarke01} found that even low levels of mass loss, such as would 
be expected in an EUV-only photoevaporation scenario, would nonetheless 
have a dramatic effect on the disk, provided that the mass loss was concentrated 
at relatively small radii (as indeed occurs for EUV photoevaporation; 
Figure~\ref{fig:4_photoevaporation}). In that case the disk evolves 
viscously until the accretion rate $\dot{M} \sim \dot{M}_{\rm wind}$, at 
which point the wind opens up a gap in the disk near $r_g$. Subsequently 
the inner disk, now deprived of resupply from large radii, drains viscously 
onto the star, followed by a rapid dispersal of the outer disk by direct 
irradiation of the inner rim \citep{alexander06b}. The resulting ``two time scale" 
behavior -- a slow period of evolution on the viscous time of the outer 
disk, followed by rapid dispersal on a time scale set by the inner disk -- 
is in general accord with observations of prompt disk clearing \citep{wolk96}. 
Qualitatively similar 
evolution can be seen in more recent calculations that include X-ray and 
FUV contributions \citep{owen10,gorti09b}, although the generally 
higher mass loss rates imply a shift of the start of the dispersal phase 
to higher accretion rates and larger disk masses. 

The most provocative finding of the studies of \citet{gorti09} and 
\citet{owen10} is unquestionably the large disk-integrated mass 
loss rates. Since much of the mass loss is inferred to occur from 
large disk radii ($r \gtrsim 50 \ {\rm AU}$), an immediate implication 
is that the structure of isolated disks on these scales may often be determined 
largely by mass loss rather than by angular momentum transport 
\citep[similar to the case of externally irradiated systems;][]{adams04}. 
There are also interesting predicted couplings between the mass loss 
rate and other observable properties of the system. Much of the FUV 
flux seen by the outer disk arises from the accretion of inner disk 
material onto the star, so in FUV-dominated photoevaporation there 
is an indirect coupling between the inner and outer disk \citep{matsuyama03}. 
If, instead, it is X-rays that dominate photoevaporative mass loss, then 
the linear dependence of $\dot{M}_{\rm wind}$ on $L_X$ 
\citep{owen10b} hints that the stellar X-ray 
luminosity function could be the fundamental determinant of disk 
evolution around low-mass stars \citep[rather than, say, dispersion 
in the initial disk mass or angular momentum;][]{armitage03}. In 
this view, the observed anti-correlation between accretion and 
stellar X-ray luminosity \citep{neuhaeuser95} would derive from 
the suppression of accretion, and reduction in disk lifetime, for 
X-ray bright stars that drive particularly powerful winds \citep{drake09,owen10b}. 
Although this is a plausible coupling \citep[and to some degree inescapable, if X-ray 
flux determines $\dot{M}_{\rm wind}$ to the extent implied by the 
results of][]{owen10}, the modeling is still at an early stage.  
Examples of possibly confounding effects include time evolution 
of the stellar X-ray spectrum, which may be affected by disk evolution via the influence of the 
disk on stellar rotation \citep{matt10}, and variations in wind loss 
rates due to dust settling and growth.

\section{INSTABILITIES AND OUTBURSTS}
There is both direct and statistical evidence that disk accretion onto young stars 
is episodic, particularly during the Class~1 or ``protostellar" phase. The 
direct evidence comes from observations of outbursts, lasting from years to 
decades, in FU~Orionis variables \citep{hartmann96}. The peak 
accretion rates during FU~Orionis events are high enough -- of the order 
of $\dot{M} \sim 10^{-4} \ M_\odot \ {\rm yr}^{-1}$ -- that if a typical star 
were to suffer a handful of such events, the total mass accreted would be 
comparable to estimated disk masses. That this might be the case is supported by statistical arguments, 
which show that the typical bolometric luminosity of protostars is smaller than 
would be expected from the infall rates and duration of the protostellar 
phase \citep{kenyon90}. This ``luminosity problem" for protostars 
could be solved if the typical star accretes around 50\% of its mass 
during less than 10\% of its lifetime \citep{evans09}. In this Section we 
will review candidate mechanisms for generating variability, where by 
``variability" we really mean large-scale episodes of enhanced accretion 
violent enough to explain FU~Orionis events and the luminosity problem. 
CTTs also exhibit a wealth of lower amplitude variability, likely attributable 
to the combined effects of fluctuations in $\dot{M}$ and to variable 
obscuration, which if it could be characterized well-enough (via long 
duration, uninterrupted monitoring) might ultimately be compared against 
simulations of turbulent accretion. 

Models for FU~Orionis outbursts generally invoke some 
combination of the following effects,
\begin{enumerate}
\item
The rapid change in opacity near $T \sim 10^4 \ {\rm K}$, associated 
with the ionization of hydrogen.
\item
A large change in MRI angular momentum transport efficiency at $T \sim 10^3 \ {\rm K}$, 
associated with the collisional ionization of alkali metals (\S\ref{sec:3_intro}).
\item
Gravitational instability, leading either to clump formation in the outer disk 
(followed by radial migration) or to heating in the inner disk.
\end{enumerate}
That aspects of this physics are involved in FU~Orionis is strongly indicated 
by observations; the inferred inner disk temperatures certainly exceed 
$10^3$~K or even $10^4$~K, and protostellar disks in general are 
plausibly massive enough for self-gravity to be in play \citep{eisner05}. 

\begin{figure}[t!]
\includegraphics[width=\textwidth]{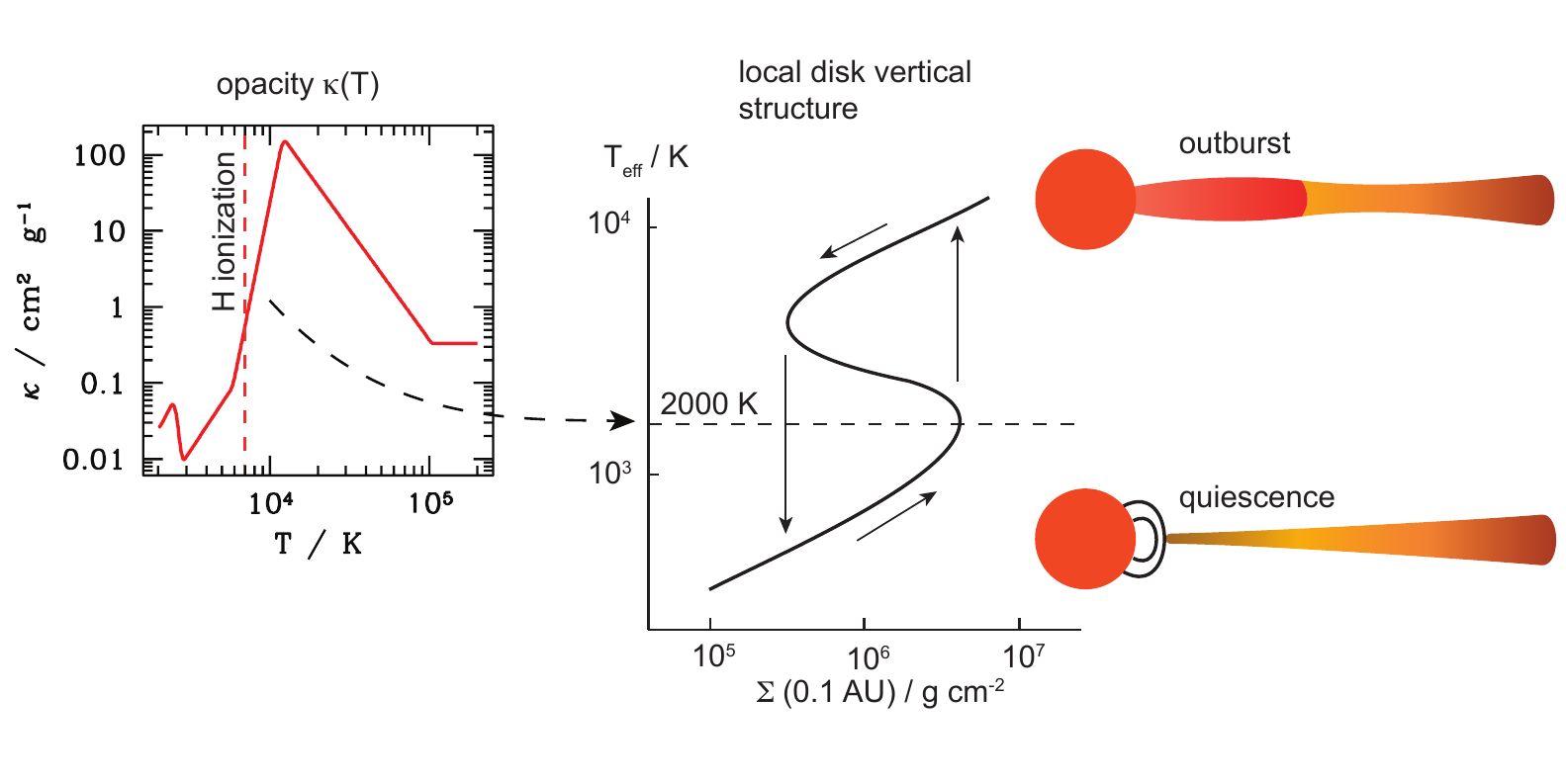}
\caption{A classical disk thermal instability is a possibility if the accretion rate 
through the inner disk is such that that the mid-plane temperature approaches 
$T \sim 10^4 \ {\rm K}$, at which point hydrogen becomes ionized and the 
opacity $\kappa (T)$ rises rapidly. Under these conditions there can be 
multiple solutions (an ``S-curve") for the local equilibrium vertical structure 
at fixed $r$ and $\Sigma$, 
with the stable possibilities being a quiescent state of low accretion rate, and 
an outburst state with a much higher temperature and accretion rate. Provided 
that these states are sufficiently well-separated, it is then possible to set up a 
global limit cycle in which the inner disk as a whole cycles between quiescent and 
outburst behavior.}
\label{fig:6_thermal}
\end{figure}

Thermal instabilities have been thoroughly studied as a potential mechanism 
for FU~Orionis outbursts \citep{bell94}. In equilibrium, the local heating rate 
$Q_+$ of a viscous disk must balance the local cooling rate $Q_-$. If that 
equilibrium is unstable to small perturbations in the central temperature $T_c$, 
i.e. if,
\begin{equation}
 \frac{{\rm d} \log Q_+}{{\rm d} \log T_c} > 
 \frac{{\rm d} \log Q_-}{{\rm d} \log T_c},
\end{equation}
then the disk is described as being thermally unstable \citep{pringle76}.  These 
conditions can be met for disks whose equilibrium mid-plane temperature 
is $T_c \sim 10^4$~K, since around this temperature the ionization of 
hydrogen leads to very rapid changes in the opacity. If a disk possesses 
regions which are locally thermally unstable, it is possible (though not 
guaranteed) that it will also be unstable to the development of a 
globally organized limit cycle. The basic idea is illustrated in Figure~\ref{fig:6_thermal}. 
The disk cycles between a quiescent state, in which the mid-plane is 
primarily cool and neutral and the accretion rate is low, and an outburst 
state in which the disk is hot and rapidly accreting.

The existence of disk thermal instability is scarcely in doubt; the physics is 
simple and models based upon it are used successfully to explain dwarf nova 
outbursts \citep[e.g.][]{hameury98}. However, the simplest models, in which 
thermal instability is the sole mechanism responsible for outbursts, only 
work if angular momentum transport in the inner disk is surprisingly 
inefficient. The one-dimensional models of \citet{bell94}, which remain 
the most detailed attempt to model FU~Orionis events, can match the 
observed time scales of outbursts if $\alpha \approx 10^{-4}$ when the 
mid-plane is neutral, and $\alpha \approx 10^{-3}$ when it is ionized. 
The low value of $\alpha$ in the outburst state, in particular, is hard 
to reconcile with either theoretical expectations for the efficiency of the 
MRI, or with observational constraints derived from protoplanetary 
disk lifetimes. As a result, and notwithstanding the otherwise 
satisfactory agreement between thermal instability models and 
observations \citep{bell95}, it appears more likely than not that 
additional physical processes contribute to the FU~Orionis 
phenomenon. Two classes of ideas have been proposed. 
The first maintains the assumption that thermal instability is 
responsible for the phenomenology of outbursts, but drops the 
requirement that the inner disk be spontaneously unstable due 
to the slow axisymmetric accumulation of matter. Instead, it is 
proposed that some other process is able to dump matter 
rapidly into the inner disk, thereby triggering an outburst  
\citep{clarke90}. Potential triggers include perturbations 
from companions \citep{bonnell92,pfalzner08,forgan10b}, 
or the inspiral of clumps formed in an outer self-gravitating 
region of the disk \citep{vorobyov05,vorobyov10,boley10,nayakshin10}.

External perturbations to the inner disk almost certainly 
do occur, but it is not known whether they are strong and frequent 
enough to account for FU~Orionis events. An alternate class of 
models retains the ``self-regulated" character of the classical 
thermal instability \citep{bell94}, and solves the time scale problem 
by associating the instability with MRI physics (and specifically the 
expected change in transport properties at $T_c \sim 10^3 \ {\rm K}$) 
at larger radii, where the viscous time scale is longer. The basic 
idea is to assume that a dead zone exists at $r \sim 1 \ {\rm AU}$, 
and that the strength of any residual transport in the mid-plane 
layer is low \citep[this is an important assumption, since a significant 
residual viscosity would allow the disk to reach a steady-state;][]{terquem08}. 
Such a dead zone is massive and potentially unstable, 
since if it could once be heated above $T_c \sim 10^3 \ {\rm K}$ 
the onset of the MRI would subsequently be able to maintain a high dissipation 
state until much of the mass in the inner disk had been accreted. Self-gravity in the 
dead zone, caused by the slow accumulation of mass there, could 
provide the heating source necessary 
to realize the latent instability \citep{gammie99,armitage01}.

These early ideas of how a dead zone might give rise to outbursts 
have recently been studied in detail, using both two-dimensional 
hydrodynamic simulations \citep{zhu09} and one dimensional 
time-dependent disk models \citep{zhu10,zhu10b}. \citet{zhu10b} 
find that protostellar accretion is liable to FU~Orionis type events 
for infall rates of the order of $\dot{M}_{\rm infall} \sim 10^{-6} \ 
M_\odot \ {\rm yr}^{-1}$, with dead zone instabilities being triggered 
at radii between 1~AU and 10~AU. As noted above the mechanism is 
closely related to the classical disk thermal instability, with the main 
distinction being that the bistable structure of the disk derives from a 
rapid change in $\alpha$ with $T_c$ due to MRI physics, rather than 
from a rapid change in $\kappa$ as hydrogen is ionized. The 
models appear to be consistent with the main observed properties of 
FU~Orionis events, with the most distinctive prediction being the 
survival of a relatively massive belt of gas in the dead zone region 
after outbursts have ceased \citep{armitage01,zhu10}.  This may 
be observable with high angular resolution observations with the 
{\em EVLA} or {\em ALMA} \citep{zhu10}.

\section{SUMMARY AND OUTLOOK}
On a optimistic reading, work to date has succeeded in identifying the 
basic physical processes responsible for protoplanetary disk evolution. 
The important angular momentum processes are the magnetorotational 
instability and disk self-gravity, with hydrodynamic instabilities associated 
with radial entropy gradients also possibly contributing. None of these 
processes behaves like an $\alpha$ viscosity, but we have at least an 
outline of how their angular momentum transport efficiencies scale 
with disk conditions. The ionization state (for the MRI), and the local 
cooling time scale (for self-gravity and baroclinic instabilities) are 
critical. Photoevaporation is almost certainly able to disperse disks, 
and may rival angular momentum transport for primacy in determining 
disk evolution, although the relative importance of FUV, EUV and 
X-ray photons remains subject to debate. Simplified 
models that combine these processes are broadly consistent with 
observations of disk populations (lifetimes, accretion rates and so 
on), and may be able to account for the eruptive behavior that 
appears to be common during the protostellar accretion phase.

There are abundant reasons to suspect that this accounting of 
important physics is incomplete. Most obviously, for want of 
space, we have restricted ourselves to considering viscous disk 
models, and have ignored the alternate possibility that magnetic 
winds or magnetic braking are responsible for significant 
disk evolution. Even if the above sketch is basically correct, 
however, there is much to be done. None of the important angular 
momentum transport processes is understood at a level adequate to 
securely predict properties of disks -- such as the amplitude of 
turbulent velocities or the depth of gaps carved by planets -- that 
may be observable in the near future. Of the many open theoretical 
questions, three stand out as 
being particularly pressing,
\begin{enumerate}
\item
What is the non-linear outcome of the MRI across the full range 
of non-ideal conditions prevalent in protoplanetary disks? There 
has been a dramatic recent shift in our understanding of how the MRI 
behaves in (near)-ideal MHD, and there is a need to revisit the 
influence of Ohmic and ambipolar diffusion, and the Hall effect, 
in light of those results and new numerical techniques. The 
importance of global effects, over long time scales, remains poorly known.
\item
How does the inclusion of magnetic fields affect the fragmentation 
of self-gravitating disks, and the efficiency of angular momentum transport 
in the regime where fragmentation is avoided? Preliminary work \citep{fromang04} 
suggests that significant changes to hydrodynamic results are 
likely when magnetic fields are included, which could be important 
for the dynamics of disks and the formation of sub-stellar objects 
at large radii.
\item
How do turbulent disks interact with planets? Many processes known to 
be of fundamental importance for planet formation -- including migration, 
gap formation, and resonant capture -- involve disks, and their detailed 
study probably requires explicit specification of the nature of disk turbulence.
\end{enumerate}
The last of these is emblematic of a more general point; in protoplanetary 
disks the strength and nature of turbulence is of central importance quite 
independent of its role (if any) in disk dynamics. The effects of MHD or 
hydrodynamic turbulence on mixing, on the formation or dynamics of 
planetesimals, or even on the thermal processing of solids, are of 
at least equal interest for planet formation.

The theoretical and computational prerequisites for tackling 
problems such as those listed above are already in place, and rapid 
progress is possible. Alas, it requires not merely 
optimism, but irrational exuberance, to believe that even the successful execution of 
such a theoretical program will bring us close to a confident understanding 
of how disks evolve. Everything we know about disks suggests that no 
single, simple process dominates their dynamics. Rather, multiple 
transport and mass loss processes -- each complex in its own 
right, and coupled to other problems in star formation and disk physics -- 
contribute at different times and radii. Divining how disks 
evolve, and hence what can reasonably be inferred as to the initial 
conditions for planet formation, will thus require a much closer 
interplay of observations and theory than has been achieved to 
date. Qualitative advances in understanding could be triggered by 
new observations in any of several areas, including,
\begin{enumerate}
\item
Direct measurements of, or stringent limits on, turbulent velocity 
fields within disks. Constraints on the strength and structure of 
magnetic fields  would be just as valuable.
\item
Detection of non-axisymmetric structure. This is only expected 
at large radii, in disks that are massive and self-gravitating, 
though more speculative possibilities such as significant 
disk eccentricity cannot be ruled out.
\item
Statistical studies of disks that are based on gas, rather than 
dust, tracers. 
\item
A direct measure of the wind mass loss rate, and velocity structure, 
for a sample of disks undergoing photoevaporation driven by the 
central star.
\item
Any constraint on the structure of disks at the most uncertain, and 
important (for planet formation), scale, between 1~AU and 10~AU. 
This region is where a dead zone may (or may not) exist, and 
where the roadblock to ab initio theory, posed by the coupling of 
MHD processes to dust physics, currently appears most intractable. 
\end{enumerate}
Encouragingly, limited observational results are already available 
for most of this wish-list, along with enticing projections of what 
{\em ALMA} and other future facilities ought to be able to achieve \citep[e.g.][]{cossins10b,wolf05}. 
The resolution and sensitivity of {\em ALMA} will also surely uncover 
entirely unexpected features of protoplanetary disks, that we hope 
will point the way toward a better understanding of the dynamics of 
these objects.

\section*{ACKNOWLEDGMENTS}
I thank my colleagues at 
the Isaac Newton Institute for Mathematical Sciences at Cambridge University, 
the Racah Institute of Physics at the Hebrew University, the Kavli Institute 
for Theoretical Physics at the University of Santa Barbara, and the Kavli Institute 
of Astronomy and Astrophysics at Peking University, for hospitality and advice 
on many of the topics discussed here. Elena Rossi, Richard Alexander, Kris Beckwith, Aaron Boley, 
Kees Dullemond, Uma Gorti, Geoffroy Lesur, 
Giuseppe Lodato, Gordon Ogilvie, Ken Rice and Zhoahuan Zhu all 
provided invaluable assistance toward the completion of the review. 
My work has been supported, in part, 
by the National Science Foundation, by NASA, and by the University of Colorado's 
Council on Research and Creative Work.

\bibliography{araa}

\end{document}